\documentclass[a4paper,fleqn,usenatbib]{mnras}

\usepackage{mathptmx}

\usepackage[T1]{fontenc}
\usepackage{ae,aecompl}

\usepackage{graphicx}	
\usepackage{amsmath}	
\usepackage{amssymb}	
\usepackage{pdflscape}	
\usepackage{longtable}

\newcommand{\kms}{km~s$^{-1}$}

\newcommand{\fxcor}{\emph{fxcor}}

\newcommand{\hght}{\emph{hght}}

\newcommand{\snr}{\emph{SNR}}
\newcommand{\cat}{\ion{Ca}{ii} triplet}

\newcommand{\vh}{$V_\text{helio}$}
\newcommand{\pa}{\emph{PA}}

\title[Carbon star rotation in NGC 6822]{The rotation of the halo of NGC 6822 from the radial velocities of carbon stars}

\author[G. P. Thompson et al.]{
Graham P. Thompson,$^{1}$\thanks{E-mail: g.p.thompson@herts.ac.uk}
Sean G. Ryan,$^{1}$
Lisette F. Sibbons$^{1}$
\\
$^{1}$School of Physics, Astronomy and Mathematics, University of Hertfordshire, College Lane, Hatfield AL10 9AB, UK\\
}

\date{Accepted XXX. Received YYY; in original form ZZZ}

\pubyear{2015}

\begin{document}
\label{firstpage}
\pagerange{\pageref{firstpage}--\pageref{lastpage}}
\maketitle

\begin{abstract}
Using spectra taken with the AAOmega spectrograph, we measure the radial velocities of over 100 stars, many of which are intermediate age carbon stars, in the direction of the dwarf irregular galaxy NGC 6822. Kinematic analysis suggests that the carbon stars in the sample are associated with NGC 6822, and estimates of its radial velocity and galactic rotation are made from a star-by-star analysis of its carbon star population. We calculate a heliocentric radial velocity for NGC 6822 of $-51\pm3$ \kms\ and show that the population rotates with a mean rotation speed of $11.2\pm2.1$ \kms\ at a mean distance of 1.1 kpc from the galactic centre, about a rotation axis with a position angle of $26^\circ\pm13^\circ$, as projected on the sky. This is close to the rotation axis of the HI gas disk and suggests that NGC 6822 is not a polar ring galaxy, but is dynamically closer to a late type galaxy. However, the rotation axis is not aligned with the minor axis of the AGB isodensity profiles and this remains a mystery.
\end{abstract}




\newpage
\section{Introduction}\label{s1}
NGC 6822 (Barnard's galaxy) is a relatively bright and well studied dwarf irregular galaxy. It lies within the Local Group (LG), quite close to the Milky Way, and is the third nearest dwarf irregular galaxy after the Large and Small Magellanic Clouds. It was the first object to be shown to lie outside the Milky Way by Edwin Hubble \citep{Hubble1925}. It has no known close companions and lies in a region relatively devoid of other galaxies \citep{deBlok2000}. Despite its apparent isolation, there is some evidence of tidal interaction and a burst of star formation $100-200$ Myr ago (\citet{deBlok2000} and references therein).\

The location of NGC 6822 has a range of published values for right ascension ($\alpha$) and declination ($\delta$)\footnote{SIMBAD: $\alpha=19^\text{h}44^\text{m}56.199^\text{s}$, $\delta=-14^\circ47^\prime51.29^{\prime\prime}$ (J2000).\

NED: $\alpha=19^\text{h}44^\text{m}57.7^\text{s}$, $\delta=-14^\circ48^\prime12^{\prime\prime}$ (J2000).\

\citet{McConnachie2012}: $\alpha=19^\text{h}44^\text{m}56.6^\text{s}$, $\delta=-14^\circ47^\prime21^{\prime\prime}$ (J2000).}, but for the purposes of this paper we use $\alpha=19^\text{h}44^\text{m}56^\text{s}$, $\delta=-14^\circ48^\prime06^{\prime\prime}$ (J2000). This is at the centre of a 3 deg$^2$ field used by \citet{Sibbons2012, Sibbons2015} and is adopted here for consistency, since this study uses spectra drawn from their sample.\

The galactic distance is variously reported to lie between 450 to over 500 kpc. For the purposes of this paper, its distance is not important, but where we do use it, we again adopt the same value as \citet{Sibbons2012, Sibbons2015}, $490\pm40$ kpc, for consistency with that work. This is the value quoted in \citet{Mateo1998}. \citet{Sibbons2012, Sibbons2015} also quotes a distance modulus for NGC 6822 of $(m-M)_0=23.45\pm0.15\; (\approx~490\pm34$ kpc). In recent years, papers have been published which have provided new distance moduli using Cepheid variables. Many of these re-estimates have reduced the radial distance to NGC 6822 by about 10\% from the value used herein. The distance modulus derived by \citet{Gieren2006} is $(m-M)_0=23.312\pm0.021$ ($\approx460\pm4.4$ kpc), \citet{Feast2012} gives $(m-M)_0=23.40\pm0.05$\footnote{\citet{Feast2012} does not claim any precision with the error value, owing to the difficulty in measuring the true uncertainty in the value of the distance modulus.} ($\approx479\pm11$ kpc), and \citet{Rich2014} gives $(m-M)_0=23.38\pm0.02_\text{stat}\pm0.04_\text{sys}$ ($\approx474\pm13_\text{tot}$ kpc). In \citet{Whitelock2013}, an alternative to Cepheid variables is proposed using Mira variables for distance measurement. These are AGBs which vary in magnitude over periods of several hundred days. NGC 6822 is used as one example of the measurement of distance using Mira variables and after calibration of the method using Miras in the Large Magellanic Cloud, the distance module for NGC 6822 is given as $(m-M)_0 = 23.56\pm0.03$ mag ($\approx515\pm7$ kpc), somewhat greater than the other estimates.\

\begin{figure*}
\centering
\includegraphics[width=0.914\columnwidth]{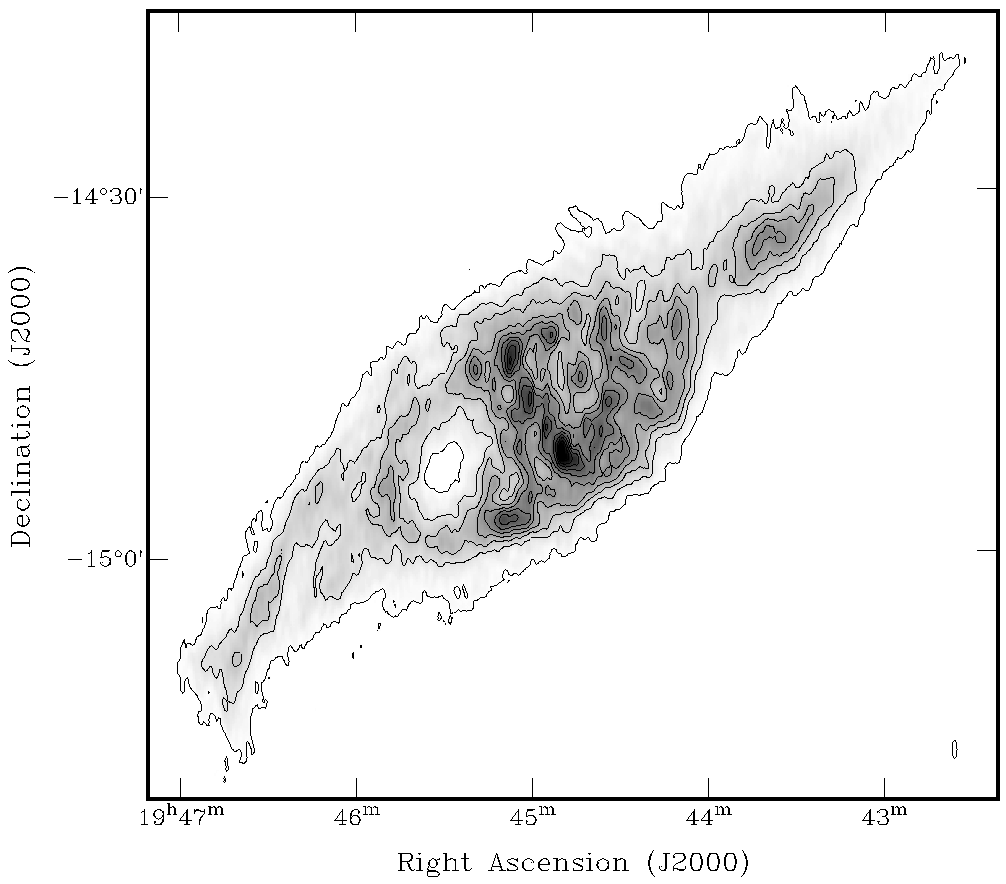}\label{deBlok2000_Fig1}
\includegraphics[width=1.026\columnwidth]{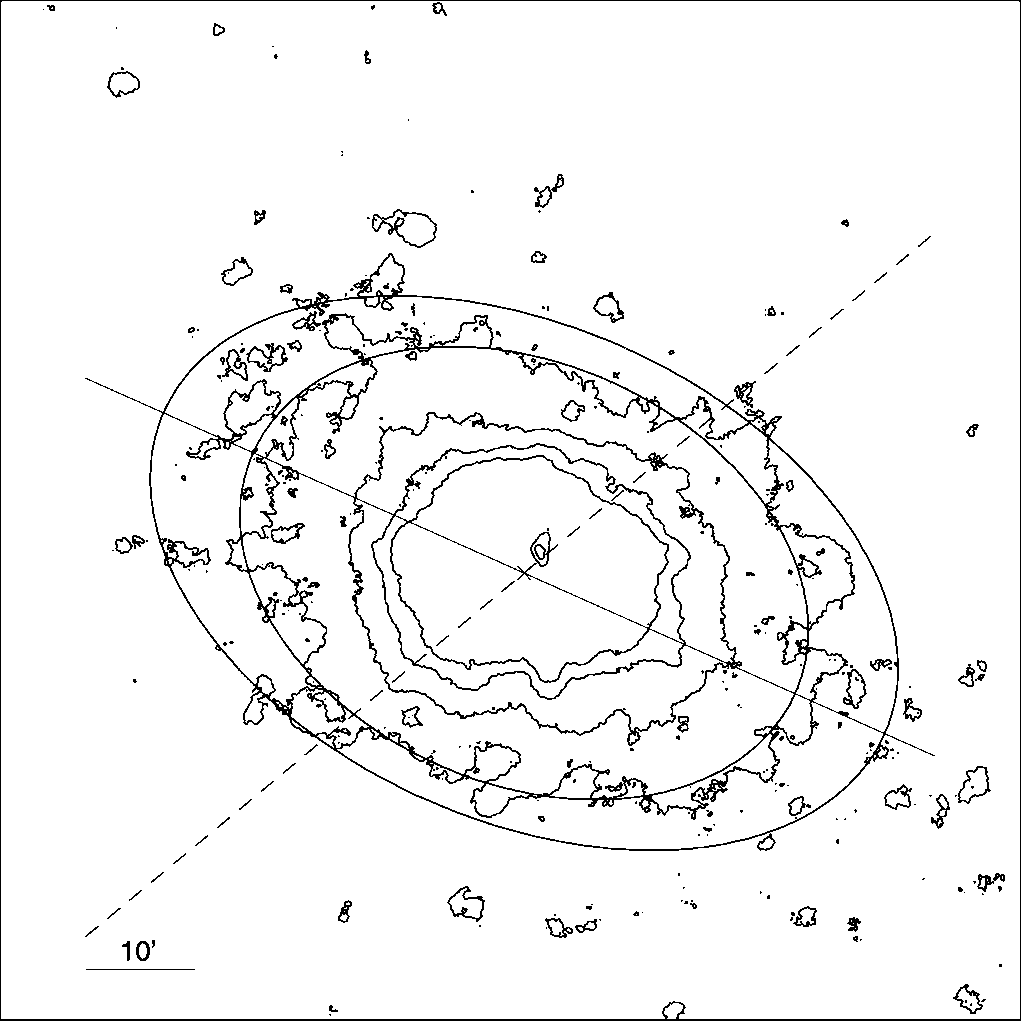}\label{Demers2006_Fig1}
\caption{Left hand panel, HI isodensity curves of NGC 6822 \citep{deBlok2000}. The HI curves exhibit a disk-like structure, which is twisted with a major axis \pa\ from $\sim110^\circ$ to $\sim130^\circ$. Right hand panel, RGB stellar isodensity profiles \citep{Demers2006}. The RGB profiles exhibit elliptical contours, where the major axis \pa\ changes from $\sim65^\circ$ (outermost contour) to $\sim80^\circ$ (innermost contour). The solid line in the right hand panel lies at $65^\circ$ and the dashed line lies $130^\circ$. The panels are shown to approximately the same scale and orientation.
}
\label{ngc6822_morphology}
\end{figure*}

Morphologically, NGC 6822 has three components: a gaseous disk of HI (Fig. \ref{ngc6822_morphology}, left hand panel), an outer spheroid of stars (Fig. \ref{ngc6822_morphology}, right hand panel) and a small central core of young stars \citep{Hodge1991}. Its HI content has been studied by \citet{deBlok2000} and its the stellar content, especially Red Giant Branch (RGB) and Asymptotic Giant Branch (AGB) stars, by \citet{Letarte2002}, \citet{Battinelli2006}, \citet{Demers2006} and \citet{Sibbons2012, Sibbons2015}. The HI gas disk appears to lie at a position angle (\pa) of $\sim130^\circ$ when observed from tip to tip \citep{deBlok2000}. However, it also appears to twist and in \citet{Weldrake2003}, the \pa\ is stated as $\sim110^\circ$, which is closer to the major axis of the central part of the disk, as shown in Fig. \ref{ngc6822_morphology} (left hand panel). \citet{Letarte2002} shows that carbon stars lie well outside the HI disc and \citet{Demers2006} and \citet{Battinelli2006} show that the isophotes of RGBs of NGC 6822 are also elliptical, where the major axis \pa\ changes from $\sim65^\circ$ (outermost contour) to $\sim80^\circ$ (innermost contour), as shown in Fig. \ref{ngc6822_morphology} (right hand panel).\

\normalsize

Much of the published velocity data for NGC 6822 has been derived from the HI gas content of NGC 6822 using the 21 cm line \citep{Koribalski2004,Weldrake2003}. \citet{Koribalski2004} measures a heliocentric radial velocity, \vh, of $-57\pm2$ \kms\ for NGC 6822 from its HIPASS spectrum, and this now a commonly cited value. \citet{Mateo1998}, and sources therein, also provide \vh\ based on the HI component, $V_{\odot,\text{radio}}$ of $-54\pm6$ \kms, and \citet{Weldrake2003} similarly provides a HI derived systemic (radial) velocity\footnote{We assume all velocities quoted in \citet{Weldrake2003} are corrected to heliocentric} in the range $-53.3$~\kms\ to $-54.7$~\kms\ at the centre of the HI gas.\

Rotation of the gas disk has also been measured at 21 cm. \citet{Mateo1998} provides a value for the rotation of the HI disk, $v_\text{rot,ISM}$, of $47\pm3$ \kms, and \citet{Weldrake2003} shows that the velocity of the HI disk ranges from $-100$ \kms\ at the NW extreme (blue-shifted) to $+10$ \kms\ at the SE extreme (red-shifted). \citet{McConnachie2012} quotes a peak observed rotation speed of $47\pm2.0$ \kms, uncorrected for inclination or asymmetric drift, citing \citet{Koribalski2004,Weldrake2003}.\

With regard to the stellar kinematics of NGC 6822, \citet{Mateo1998}, and references therein, provide a value of \vh\ based on optical measurements of $V_{\odot,\text{opt}}=-53\pm4$ \kms. \citet{Kirby2014} studies the kinematics of seven isolated dwarf galaxies in the LG from the radial velocities of red giant stars, including NGC 6822, and derives a mean heliocentric radial velocity, $\langle$\vh$\rangle$, of $-54.5\pm1.7$ \kms for NGC 6822.\

However, while there are published data for \vh, there are little regarding stellar rotation. Many publications refer to \citet{Demers2006} for stellar rotation, where the radial velocities for 110 carbon stars lying within $15'$ ($\sim2.1$ kpc) of the HI major axis, taken from a sample in \citet{Letarte2002}, are measured, although the individual stellar measurements are not published. A variation in residual velocity along the major axis of the outermost RGB isophote is interpreted to be a signature of rotation about the minor axis. As a result, the paper suggests that the carbon stars appear to rotate about an axis roughly perpendicular to the rotation axis of the HI disk, resembling a Polar Ring Galaxy (PRG). It would be extraordinarily useful if NGC 6822 is a PRG, since they are very rare. The Sloan Polar-Ring Catalogue (SPRC), which uses images from the Sloan Digital Sky Survey, includes only 6 confirmed PRGs from a total of just 275 candidates. Discovery of a nearby PRG would present a remarkable opportunity to study these galactic types at close hand. In a separate publication \citep{Demers2007} suggest that radial velocities of individual NGC 6822 giants might give some insight into rotation of the stellar population.\

\citet{Kirby2014} checked for signs of rotation in their sample of 7 dwarf galaxies, and found that only the Pegasus dwarf irregular showed obvious signs. For NGC 6822, they refer to the findings of \citet{Demers2006}. They comment that while a rotation of the order of $\sim10$ \kms\ is possible in NGC 6822, their data were too highly obscured by velocity dispersions to be certain.\

The published studies of the galaxy's stellar rotation are complemented by other studies of the motion of globular clusters \citep{Veljanoski2015} and planetary nebulae \citep{Flores2014}. \citet{Veljanoski2015} studies the globular cluster (GC) system of NGC 6822, in which 8 GCs are presently identified (see Table 1 therein). The radial velocities of 6 of the GCs are given and $\langle$\vh$\rangle$ is deduced to be between $-59$ and $-60$ \kms. The spatial distribution of the GCs in their data is rather linear and lies approximately parallel to the major axis of the AGB isophotes. Three possible dynamical models are considered:
\begin{itemize}
\item `Disk model', in which the rotation axes are similar to that suggested by \citet{Demers2006}. The resulting rotation rate is $12\pm10$ \kms.
\item `Cigar model', where the cluster system shares the same rotation axis as the HI disk. In this case the rotation rate is determined to be $56\pm31$ \kms.
\item A scenario where there is no net rotation by disconnecting any relationship between the gas disk and the stellar component.
\end{itemize}
In none of these models are they able to independently determine the \pa\ of the rotation axis owing to the small number of objects in their sample of GCs, but their preferred case is the `Disk model', owing to its low rotation rate. They argue that in the case of the `Cigar model' the rotation is too high, and would cause the sample to flatten into a disk which they do not observe.\

In \citet{Flores2014}, the motion of 10 planetary nebulae (PNe) in the galaxy are studied. PNe are the evolutionary stage reached by AGBs after the end of their lives, and may thus provide complementary kinematic information of the galaxy. \citet{Flores2014} compute the mean $\langle\text{\vh}\rangle$ of the PNe to be $-57.8$ \kms. They compare their result with four extended star clusters (ESGs) which have a mean \vh\ of $-88.3$ \kms\ \citep{Hwang2014} and with a C-star mean \vh\ of $-32.9$ \kms\ derived by \citet{Hwang2014}. The ESGs lie $\sim10.5$ kpc from the centre of NGC 6822 and show no signs of rotation \citep{Hwang2014}. \citet{Flores2014} infer that the PNe, C-stars and clusters belong to different dynamical systems, but with so few PNe, they are unable to reliably fit the radial velocity data to either system.\

This paper provides the results of a new study of the kinematics of NGC 6822, based on the spectra of individual stars obtained in 2011 by \citet{Sibbons2015}, and follows studies by \citet{Sibbons2012, Sibbons2015} which classify stars, photometrically and spectroscopically, from a set taken from the catalogue of \citet{Letarte2002}. These studies were able to distinguish C-type and M-type AGBs from other types, so that although \citet{Demers2007} cautions that the separation of carbon stars associated with NGC 6822 from foreground Galactic dwarfs might be difficult, our present study has the advantage of using the spectra of stars which have already been classified as carbon stars in NGC 6822, with a high degree of confidence.\

The radial velocities of well over 100 stars within a 4~kpc radius of the centre of NGC 6822, (the {\it inner region}), and well over 100 stars outside the 4~kpc radius, (the {\it outer region}), are measured and reported. In the inner region many of these stars are classified as carbon stars and are expected to be associated with the galaxy \citep{Sibbons2015}. In the outer region, most of the stars are expected to be foreground Milky Way stars \citep{Sibbons2012,Sibbons2015}.\

Based on the aggregated sample of carbon stars in the inner region, a value for the heliocentric radial velocity of NGC 6822 is derived. The kinematics of individual stars in the sample indicate that they belong to the same population, supporting \citet{Sibbons2015}. Moreover, the radial velocity measurements are accurate enough to reveal rotation of the stellar component about an axis through the centre of the galaxy, and to derive its rotational speed and a new \pa\ for the axis of rotation. In \S \ref{S2}, we discuss the data set, its origins and its limitations. In \S \ref{S3}, we discuss the method used to analyse the data set including the selection of objects for analysis. In \S \ref{S4}, we discuss the results and in \S \ref{S5}, we draw conclusions.

\section{Observations and Data}\label{S2}
This study uses AGB stars as tracers of kinematic properties of the galaxy. AGB stars are amongst the brightest stars in an intermediate-age or old stellar population and can be resolved in galaxies beyond the Milky Way, making it possible to study them individually. AGBs tend to divide into those with oxygen rich atmospheres (M-types), and those with carbon rich atmospheres (C-types or carbon stars). They have relatively low surface temperatures, $T_\text{eff}$, ranging from as low as 2200 K \citep{Matthews2015} for old stars near the end of their lives up to $\sim4000$ K for intermediate age stars\footnote{In \citet{vanBelle2013}, 12 carbon stars are found to be in a range of $T_\text{eff}=2381\pm81$ K to $3884\pm161$ K, with a mean at $2800\pm270$ K.}. The continuum peaks at wavelengths in the near infra red, with increasing brightness through the $I,\; J,\; H$ and $K$ bands.\

The spectrum of a typical AGB contains a number of absorption features, predominantly broad TiO bands in M-type AGBs, broad CN bands in C-type AGBs. The $I-$band spectral region also contains the \cat\ lines in both types of AGB. The \cat\ lines are ideal for measuring radial velocities as they are narrow atomic features and lie in a position where the continuum is relatively strong. In this study, we use the spectra of carbon rich AGBs classified by \citet{Sibbons2015} as follows: confirmed C-type stars (C); tentative C-type stars (C:), confirmed C-type stars which exhibit H$\alpha$ and sometimes [S II] and [N II] emission (Ce) and tentative C-type stars which similarly exhibit emission (Ce:).\

\begin{figure}
\centering
\includegraphics[width=\columnwidth]{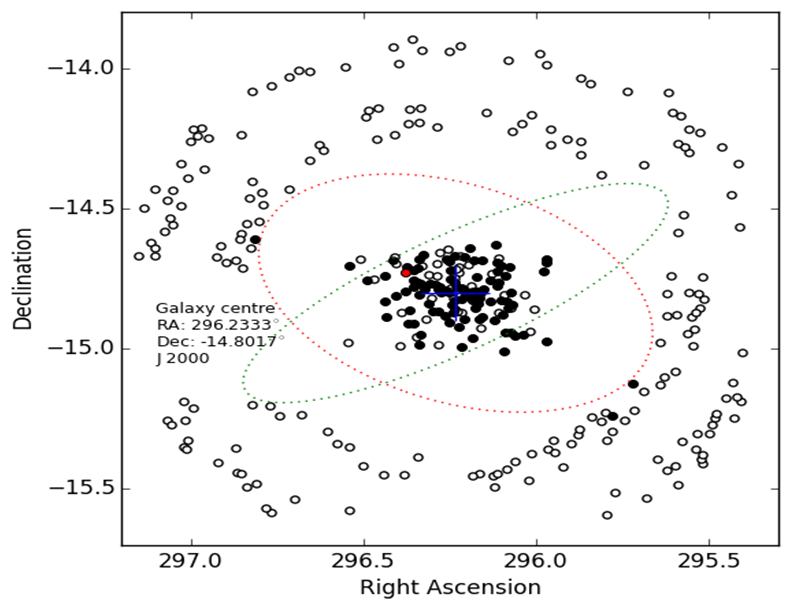}
\caption{The locations of 323 NGC 6822 objects targeted by \citet{Sibbons2015}, shown over $1.8^\circ$ of the sky. The black points and red point show 99 objects classified by \citet{Sibbons2015} as C-type AGBs. The red point marks object \#31, which is discussed in the text. The 3 outermost C-type AGBs lie more than 4 kpc from the galactic centre and were not used in the analysis. The open circles mark the remaining objects which are given other classifications. The dotted ellipses approximate the location and extent of the gas disk and outermost stellar isodensity profile shown in Fig. \ref{ngc6822_morphology}. The concentric appearance of the outer objects is due to the layout of the fibres on the spectrograph field plate.}
\label{ngc6822_objects}
\end{figure}

Wide band, low resolution spectra for 323 target objects, out to $\approx8$ kpc from the galactic centre, were obtained using the AAOmega multi-fibre spectrograph on 30 and 31 August 2011 \citep{Sibbons2015}. An optical fibre was placed over each target object position on the field plate of the spectrograph and throughout this paper, objects are referred to by their fibre number. Wide band gratings 385R and 580V were used in the spectroscopy but this study uses only the spectra from grating 385R, which has a bandpass from 5687 \AA\ to 8856 \AA, covering the \cat, dispersed at 1.6~\AA\ pixel$^{-1}$. The spectra were reduced to science frames, using the AAOmega pipeline. The spatial position of each object is plotted in Fig. \ref{ngc6822_objects}, which shows 135 objects in the inner region, and 188 objects in the outer region. Ninety eight black points and one red point show the positions of the C-type AGBs. The red point is the location of object \#31 which was utilised in the processing of the data described later. Three black points located just outside the outermost isodensity profile were not used in the analysis, so the total number of carbon stars used is 96. Ellipses which reproduce the location and extent of the gas disk and outermost RGB isodensity profile of NGC 6822 are also plotted on Fig.~\ref{ngc6822_objects}.\

\normalsize

\begin{figure*}
\centering
\includegraphics[width=\columnwidth]{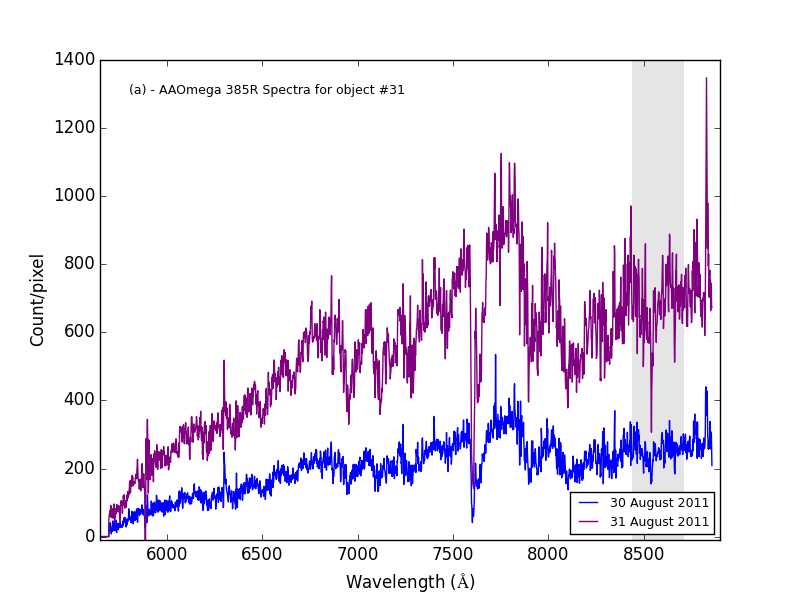}\label{T31Spectrum_full}
\includegraphics[width=\columnwidth]{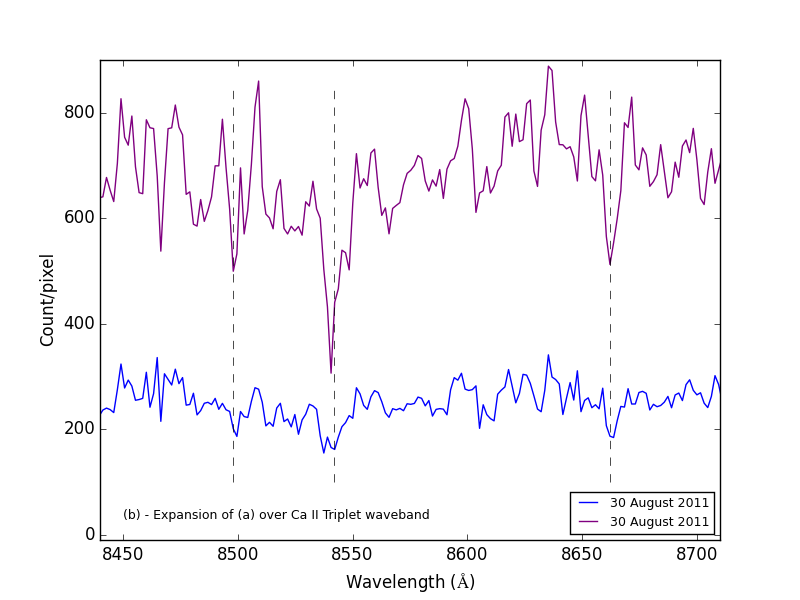}\label{T31Spectrum}
\caption{Spectra for Object \#31. Left hand panel: Full red spectrum from AAOmega grating 385R for object \# 31 on 30 and 31 August 2011. Object \#31 is spectrally classified as a C-type star in \citet{Sibbons2015} and has CN absorption bands at 5730, 6206, 6332, 6478, 6631, 6925, 7088, 7259, 7437, 7876, 8026 \AA\ which are characteristic of this type of star. The large absorption feature at 7594 \AA\ is due to telluric O$_2$. The grey band indicates the waveband 8440\AA\ to 8710\AA\ containing the \cat\ features. Right hand Panel: The same spectra expanded over the waveband 8440\AA\ to 8710\AA\ containing the \cat\ features used for radial velocity measurement. The centres of the \cat\ absorption lines are indicated by the vertical dashed lines at 8498.02~\AA, 8542.09~\AA, 8662.14~\AA.
}%
\label{Typical_Spectra}
\end{figure*}

Fig. \ref{Typical_Spectra} shows the spectra of a typical C-type AGB, object \#31, for both observing nights. It is evident that the spectrum of 31 August has a higher count in the continuum than that of 30 August and this results in a correspondingly higher signal to noise ratio, \snr. This is generally the case for all spectra and led us to place more weight on the results for the night of 31 August. The waveband containing the \cat\ is shaded in grey in the left hand panel and its absorption features indicated at 8498.02~\AA, 8542.09~\AA\ and 8662.14~\AA\ (rest) in the right hand panel.\

\normalsize

Although the velocity resolution of the grating was rather low, $\approx 56$~\kms\ pixel$^{-1}$ in the \cat\ waveband of interest, and the individual errors rather higher, there were enough stars present in the sample to reduce the standard error in the mean velocity to $\sim8$\%.\

\section{Radial Velocity Measurements}\label{S3}

To measure relative velocities, we used the cross-correlation technique of \citet{Tonry1979}, as implemented in the IRAF\footnote{IRAF is distributed by the National Optical Astronomy Observatory, which is operated by the Association of Universities for Research in Astronomy (AURA) under a cooperative agreement with the National Science Foundation.} routine \fxcor, star by star. We produced two composite spectra from the spectra of the AGBs in the sample, which were used as the template required by this technique. One template was made from C-type AGBs from the inner region, and the second from the dM-type stars in the outer region as classified by \citet{Sibbons2015}. Both templates were corrected to the `rest frame', and then used to measure the radial velocities of the stars. The C-type template was used for all stars, while the M-type template was used for the objects in the outer region only.\

\begin{figure}
\centering
\includegraphics[width=\columnwidth]{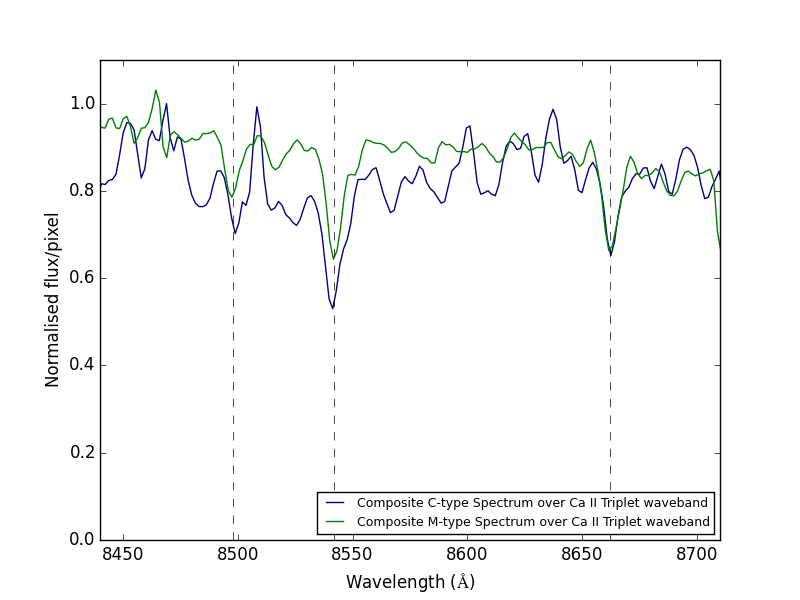}
\caption{Composite template spectra in the rest frame used for the radial velocity measurements. The centres of \cat\ features are shown by the vertical dashed lines.
}
\label{sgi31167_Composite_Template.png_Composite3_ArcTrunc_xcorr.png}
\end{figure}

The use of a composite template ensures that its spectrum is well matched to the object spectra, that is, it includes the same instrument and observing systematics as the object spectra. Moreover, the combination process makes a significant reduction in the noise content in the template, an important consideration as our spectra had originally been taken with a relatively short exposure time. The composite template spectra are shown in Fig. \ref{sgi31167_Composite_Template.png_Composite3_ArcTrunc_xcorr.png}. The undulations in the C-type spectrum are due to the remaining \cat\ and to CN absorption lines.\

\normalsize

\fxcor\ generates a cross-correlation function of an object spectrum and the template spectrum, and fits a Gaussian to it. As the template is in the rest frame, the position of the Gaussian peak with respect to the baseline determines the radial velocity, $V_r$, of the object, and application of the heliocentric correction\footnote{$-19.5$~\kms\ (30 August 2011) and $-19.8$~\kms\ (31 August 2011)} yields an object's heliocentric radial velocity, \vh. Initial experimentation and visual inspection were performed to find the best parameters to use in \fxcor\ to minimise errors and maximise the number of successful cross-correlation operations. It was found that a 9 pixel Gaussian fit ($\equiv$ 13.5 \AA\ or 717 \kms) fitted the shape of the peak of the cross-correlation function closely and did not limit the number of successful cross-correlations. All spectra were truncated to the waveband 8440 \AA\ to 8710 \AA, which covers the \cat, to ensure that IRAF calculated the velocity resolution appropriate to this waveband, and the cross-correlation bandwidth was restricted to 8450 \AA\ to 8700 \AA\ to reduce spectral noise coming from outside this waveband.\

The quality of the spectra is important. As well as low spectral resolution, many of our spectra had low \snr\ ($<10$ pixel$^{-1}$), causing noisy cross-correlation functions. These sometimes returned unreliable results or failed to give a result at all. To overcome these difficulties, selection criteria were adopted to rule out any spectra which might give rise to unreliable results. The criteria were based on the strength of the cross-correlation function, \hght\ and on the radial velocities returned by \fxcor\ as described below.\

It was found, by inspection, that cross-correlation functions with \hght\ $>0.4$ give a strong peak with low sidebands, but those with \hght~$<0.2$ do not. In these cases, the peaks are weak with respect to the sidebands and in some cases, multiple peaks occur. Between these limits, the quality of the cross-correlations degrades but useful results can be achieved in many cases. After visual inspection of each cross-correlation function, we excluded all cross-correlation functions where \hght $<0.2$.\

We then examined the \fxcor\ velocity returns. In a number of cases, unrealistically high values of \vh\ were observed. These can come about by selection of the `wrong' peak by \fxcor\ in noisy or multi-peak cross-correlation functions. Values of \vh\ as large as $\pm1554$~\kms, $\pm4184$~\kms\ or $\pm5734$~\kms\ arise, which are regarded as unrealistic, and such values tend to be associated with \hght\ values close to or less than 0.2. The distribution of velocities returned from \fxcor, corrected to \vh, was analysed and showed that most velocities fall within $\pm200$~\kms. All cross-correlation functions where \hght $>0.4$ yielded velocities within this range so, on this basis, heliocentric velocities falling outside this range were excluded.\

\begin{landscape}
\begin{table}
\caption{The first 18 lines of the individual results for objects meeting the criteria (\hght$\geqslant 0.2$, $-200$ \kms\ $\leqslant V_\text{helio} \leqslant +200$ \kms). The full table can be found in Appendix \ref{App_A} to the electronic version of this paper}
\label{Results}
\centering\scriptsize
\begin{tabular}{c c c c c c c c | c c c c | c c c c | c}
\hline
\multicolumn{17}{c}{Objects meeting the criteria (\hght$\geqslant 0.2$, $-200$ \kms\ $\leqslant V_\text{helio} \leqslant +200$ \kms).}\\ 
\hline\hline
[1]&[2]&[3]&[4]&[5]&[6]&[7]&[8]&[9]&[10]&[11]&[12]&[13]&[14]&[15]&[16]&[17]\\
&&&&&&&&\multicolumn{4}{c|}{30 August 2011}&\multicolumn{4}{c|}{31 August 2011}&\\ 
\#Fibre & ID & RA & Dec & $I-$band & distance & phot. & spec. & \snr & \hght & \vh & $V_\text{err}$ & \snr & \hght & \vh & $V_\text{err}$&Flag\\
& &$(^\circ)$ & $(^\circ)$ & (mag.) & (kpc) & class. & class. & & & (\kms) & (\kms) & & & (\kms) & (\kms)&\\
\hline
\hline
1&217716&296.23&-14.86&19.3&0.52&M&Ce:&7&0.39&-10&49&9&0.48&-35&36&C\\
2&239630&296.33&-14.95&19.4&1.52&C*&C:&6&0.43&-55&41&7&0.54&-55&27&C\\
3&210316&296.43&-14.83&18.9&1.69&M&C:&5&0.59&15&25&6&0.76&-11&16&C\\
4&174035&296.37&-14.71&19.1&1.4&M*&C&3&0.37&-26&57&5&0.27&-43&67&C\\
6&199974&296.37&-14.8&19.2&1.17&M&C&...&...&...&...&9&0.65&4&24&C\\
7&194949&296.35&-14.78&18.9&1&C*&C:&12&0.72&-1&34&18&0.75&-14&23&C\\
8&242030&296.35&-14.96&18.8&1.64&...&...&6&0.6&-53&21&9&0.7&-62&22&...\\
9&220271&296.39&-14.87&...&1.43&...&...&5&0.48&-28&40&8&0.59&-63&27&...\\
10&188246&296.49&-14.76&18.3&2.12&M&C:&6&0.57&-40&29&10&0.8&-39&16&C\\
11&172656&296.54&-14.71&19.3&2.66&C*&C&5&0.29&-125&69&8&0.39&-70&41&C\\
12&211898&296.31&-14.84&18.8&0.7&M&C&9&0.57&-31&26&7&0.74&-21&20&C\\
14&246838&296.54&-14.98&...&2.99&...&...&4&0.36&-78&37&5&0.57&-61&28&...\\
15&206867&296.3&-14.82&...&0.59&...&...&7&0.42&1&58&8&0.42&-3&38&...\\
17&225812&296.32&-14.9&...&1.06&...&...&4&0.33&0&75&6&0.51&-10&31&...\\
18&190283&296.34&-14.76&19.6&0.96&M&C&6&0.45&-44&33&5&0.36&-51&36&C\\
19&168284&296.41&-14.69&19.0&1.78&C*&C:&9&0.48&40&42&10&0.48&21&34&C\\
21&213379&296.24&-14.85&19.9&0.39&C*&Ce:&4&0.4&-25&42&...&...&...&...&C\\
22&200573&296.27&-14.8&19.0&0.34&M&Ce&8&0.37&10&73&9&0.72&-12&17&C\\
\hline
\end{tabular}
\end{table}\normalsize

\begin{table}
\centering\scriptsize
\begin{tabular}{p{3cm}p{20cm}}\\
Column&\\
1 \& 2&Fibre number and the target identifier respectively as defined by \citet{Sibbons2015}.\\
3 \& 4&Right ascension (RA) and declination (Dec) of each object (J2000) according to \citet{Sibbons2015}.\\
5&$I-$band magnitude of C-type AGBs in the inner region from the catalogue of \citet{Letarte2002}.\\
6&Distance of the object from the putative galactic centre ($\alpha=19^h:44^m:56^s$ ($296.2333^\circ$), J2000 and $\delta=-14^\circ:48':06''$ ($-14.8017^\circ$), J2000) centred on the optical co-ordinates of NGC 6822\footnote{Different catalogues and databases have differing values for the centroid of NGC 6822. These values are adopted to be consistent with \citet{Sibbons2012, Sibbons2015}} adopted by \citet{Sibbons2012, Sibbons2015} and a galactic distance of 490 kpc \citep{Mateo1998}.\\
7 \& 8&Photometric classification of each object from \citet{Sibbons2012}, and its corresponding spectroscopic classification from \citet{Sibbons2015}.\\
9 \& 13&\snr\ of each object calculated in the band 8560 \AA\ to 8650 \AA\ for the nights of 30 August 2011 and 31 August 2011 respectively.\\
10, 11 \& 12&\hght, the heliocentric radial velocity (\vh) and the velocity error ($V_\text{err}$) returned from \fxcor\ for the night of 30 August 2011.\\
14, 15 \& 16&\hght, the heliocentric radial velocity (\vh) and the velocity error ($V_\text{err}$) returned from \fxcor\ for the night of 31 August 2011.\\
17&Flag showing which objects were used in the composite templates: ``C'' for the C-type composite and ``M'' for the M-type composite.\\
\end{tabular}
\end{table}\normalsize

\end{landscape}

\section{Results and Analysis}\label{S4}
\subsection{Cross-correlation results with the C-type template}\label{s4.1}
All 323 objects were cross-correlated with the C-type composite template. Application of the criteria discussed in \S \ref{S3} resulted in 128 successful cross correlations in the inner region and 107 in the outer region on 30 August, and 129 in the inner region and 124 in the outer region on 31 August. From the population of 96 C-type AGBs in the inner region, 89 gave results meeting the selection criteria on 30 August and 92 on 31 August.\

The full results of the radial velocity measurements and their individual errors over both nights are shown in full in the Appendix to the electronic version of this paper. The first 18 lines of these results are reproduced in Table \ref{Results}, with explanations of the column contents. Spectra which failed to meet the selection criteria are not included.\

Table \ref{summary_results} summarises the results. Column [3] shows the mean heliocentric radial velocities, $\langle \text{\vh} \rangle$, of the sample. For the inner region, this can be considered a good approximation to the heliocentric radial velocity of NGC 6822. The mean individual error in column [4] is large and reflects the low resolution and low \snr\ of the spectra. $\sigma(\text{\vh})$ in Column [5], is the velocity dispersion, which is also high, nevertheless, the standard error of the mean (SE) in column [6] is small, $3-8$~\kms, implying that, provided the individual errors are random, the mean and median (column [7]) heliocentric radial velocities are well constrained. The mean velocities show reasonable consistency from night to night, with differences similar to the standard error in the mean, SE. The inner region differs by $>-24$ \kms\ from the outer region\footnote{This supercedes the preliminary data stated in \citet{Sibbons2015} and is the result of more detailed work.}, supporting the suggestion that the objects in the outer region are not part of NGC 6822.\

\begin{table*}
\caption{Summary of Results for spectra with $\text{\hght} \geq 0.2$ and $\vert \text{\vh} \vert<200$~\kms.
}
\label{summary_results}
\centering\footnotesize
\begin{tabular}{l c c c c c c}\\
Template Object&No. of&$\langle V_\text{helio}\rangle$&$\langle V_{err}\rangle$&$\sigma(V_\text{helio})$&SE&Median \vh\\
Composite Template&Objects&[\kms]&[\kms]&[\kms]&[\kms]&[\kms]\\
&[2]&[3]&[4]&[5]&[6]&[7]\\
\hline&&&&&&\\
{\bf 30August2011}&&&&&&\\
Inner region (C-template)&128&$-45$&$\pm42$&$45$&$\pm4$&$-43$\\
\\
Outer region (C-template)&107&$-20$&$\pm63$&$81$&$\pm8$&$-19$\\
Outer region (M-template)&100&$-28$&$\pm54$&$60$&$\pm6$&$-22$\\
&&&&&&\\
{\bf 31August2011}&&&&&&\\
Inner region (C-template)&129&$-51$&$\pm32$&$29$&$\pm3$&$-51$\\
\\
Outer region (C-template)&124&$-17$&$\pm60$&$66$&$\pm6$&$-9$\\
Outer region (M-template)&133&$-18$&$\pm49$&$58$&$\pm5$&$-19$\\
\hline\hline&&&&&&
\end{tabular}
\end{table*}\normalsize

Thus, we conclude that the heliocentric radial velocity of NGC 6822, as based on the sample of the inner C-type AGB population, lies in the median range of $-43\pm4$ to $-51\pm3$ \kms\ ($1 \sigma$). The data for 31 August are preferred, owing to their better \snr, lower errors and greater number of successful cross-correlations than the data for 30 August, and provides \vh $=-51\pm3$ \kms\ ($1 \sigma$). This compares favourably with published values described earlier.\

\begin{figure*}
\centering
\includegraphics[width=\columnwidth]{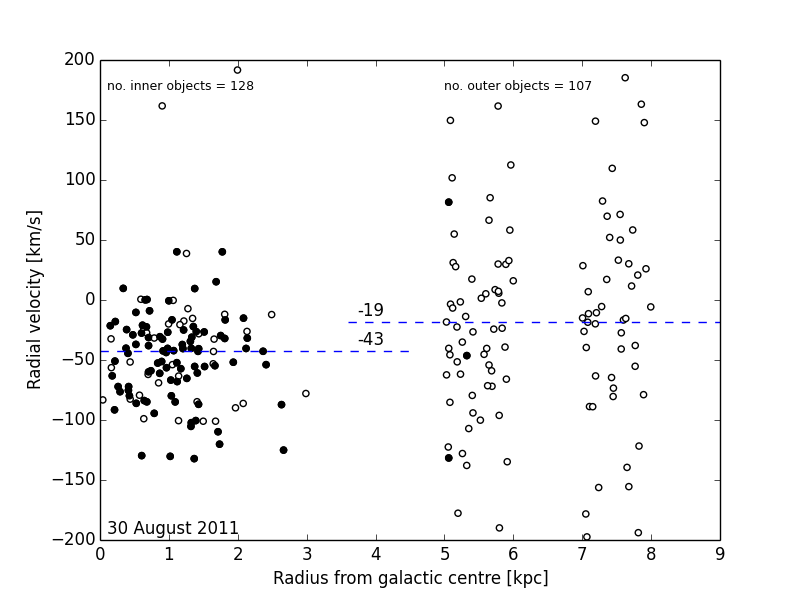}
\includegraphics[width=\columnwidth]{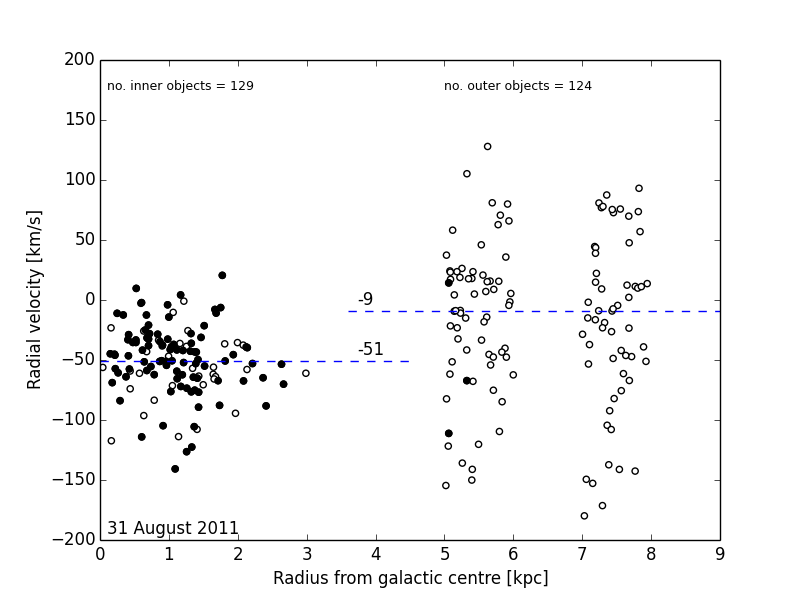}
\caption{Relationship between \vh\ and distance $D$ from the galactic centre, left panel for 30 August 2011 and right panel for 31 August 2011. The horizontal dashed lines represent the median radial velocity of the sample from Table \ref{summary_results}. C-type AGBs are shown as black points.
}
\label{vr_d_Composite}
\end{figure*}

Fig. \ref{vr_d_Composite} plots \vh\ of all objects which meet the acceptance criteria, by distance $D$ from the galactic centre, for both nights. It shows a number of differences between the inner and outer populations. The inner objects tend to clump more closely around the median value for the inner region and they are therefore likely to be part of the same dynamical system, while the outer objects are much more widely spread about a different median value, indicating a separate population. This suggests that the inner population is associated with NGC 6822, supporting similar conclusions in \citet{Sibbons2012, Sibbons2015}, while the outer population is not.\

\normalsize

The aggregated results of the cross-correlation of the spectra of objects in the outer region using the M-type template are also presented in Table \ref{summary_results}. Both the formal errors, $\langle V_\text{err} \rangle$, and the spread of results, $\sigma(\text{\vh})$, are reduced by switching to the M-type template in the outer region. The improvement observed in outer region errors when an M-star template is used supports the conclusion that the outer stars are predominantly dM-type, dK-type and other unclassified foreground stars of the Milky Way, \citep{Sibbons2015}.\

\subsection{Rotation of the carbon star population}\label{s4.4}
\begin{figure*}
\centering
\includegraphics[width=0.8\textwidth]{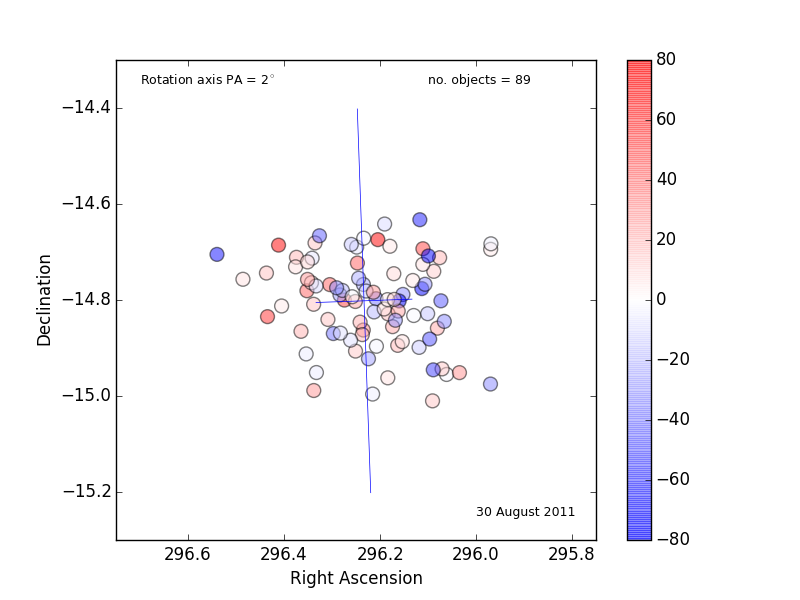}\label{rv_plot_a}
\includegraphics[width=0.8\textwidth]{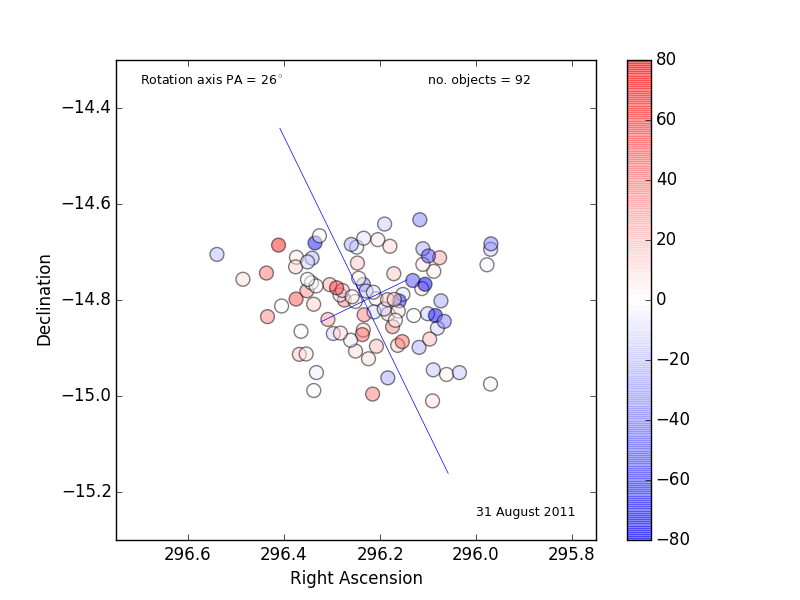}\label{rv_plot_b}
\caption{Residual velocity, $\text{\vh}-\langle \text{\vh}\rangle$, plots of the carbon star sample in the inner region of NGC6822. The long axis represents a derived axis of rotation and passes through the galaxy centre at the intersection with the short axis.
}%
\label{ngc6822_motion}
\end{figure*}

In the inner region, the \vh\ of each C-type object was compared to the sample mean, $\langle \text{\vh}\rangle$, which we take to be the mean motion of the galaxy. This provides a measure of the motion of each object in the rest frame of the galactic centre of NGC 6822. Fig. \ref{ngc6822_motion} plots the locations of the inner C-stars, colour coded by the value of its residual velocity, $\text{\vh}-\langle \text{\vh}\rangle$, using a colour range from red (for positive values) to blue (for negative values). The colour range is consistent with red and blue shift and indicates whether an object is receding or approaching in the rest frame of NGC 6822. In the lower panel for 31 August, it can be seen that receding objects dominate the south east sector and approaching objects the north west sector, suggesting rotation of the C-star population about an axis, oriented approximately NE-SW. In the upper panel, for the night of 30 August, the pattern is less obvious but still apparent.\

\normalsize

\begin{figure*}
\centering
\includegraphics[width=0.8\textwidth]{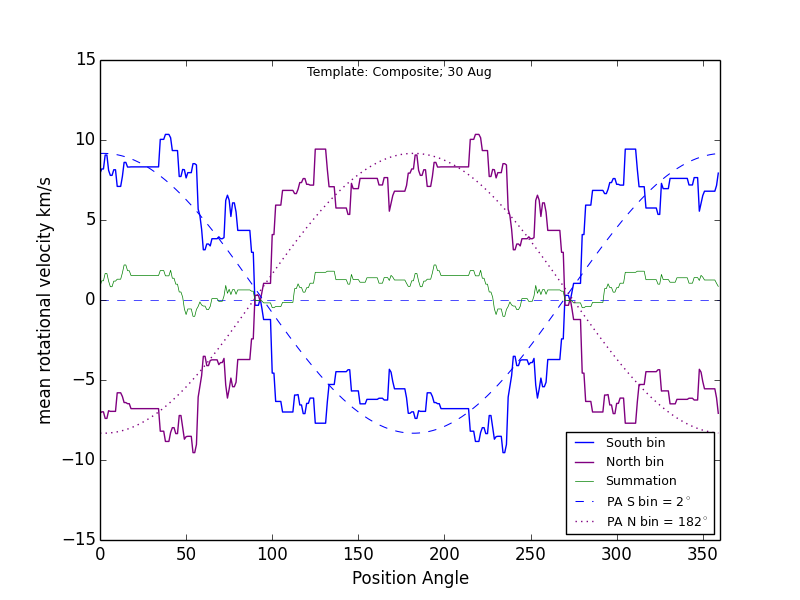}\label{rv_plot_c}
\includegraphics[width=0.8\textwidth]{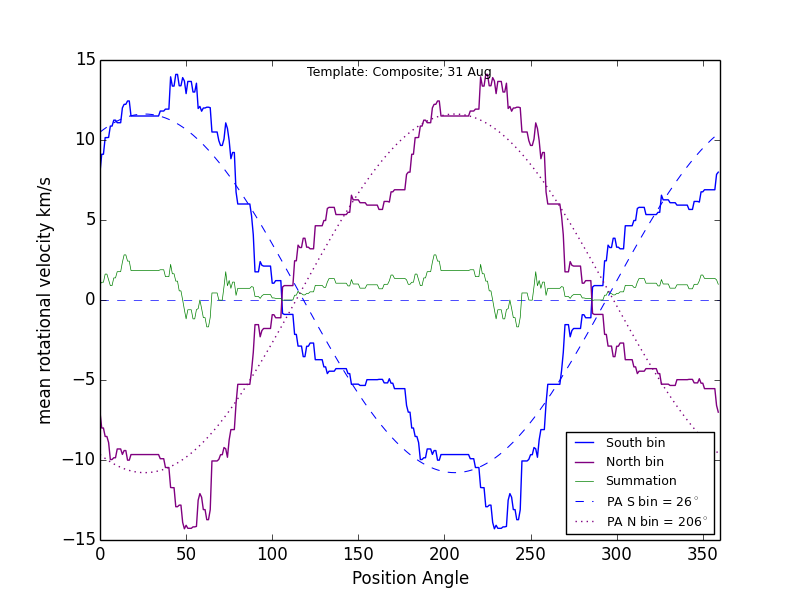}\label{rv_plot_d}
\caption{Rotational signature of carbon star component in NGC 6822 derived by plotting the mean residual velocities, $\langle \text{\vh}-\langle \text{\vh}\rangle\rangle$, on either side of the long axis as a function of its position angle.
}%
\label{ngc6822_motion_2}
\end{figure*}

To examine this further, an axis of rotation was hypothesised lying in the plane of the sky and passing through a pivot point at the centre of the galaxy. The axis was rotated through $360$ steps of $1^\circ$, starting at a position angle (\pa) of $0^\circ$. At each step, the mean of the residual velocities of the stars on either side of the axis was computed. The subsample of stars falling on one side of the line is termed the `North Bin', and for an axis with \pa $=90^\circ$, this is defined naturally. The subsample falling on the other side of the line is termed the `South Bin'. Membership of each bin changes as the axis of rotation sweeps around and the value of its mean velocity residual also changes. By plotting the mean velocity residuals over $360^\circ$, a clear rotational signature is observed, see Fig. \ref{ngc6822_motion_2}. The purple curves plot the mean rotational velocity of the `North Bin' and the blue curves plot the mean rotational velocity of the `South Bin'. In both bins and on both nights the curves rise and fall in opposition as the putative axis is rotated, confirming our impression of a rotating population.  Furthermore, the curves for the 30 August and 31 August both suggest a similar sense of speed and rotation, even though the curves for 30 August are noisier.\

\normalsize

The green curve in each case shows the sum of the average `North' and `South' velocities, which is generally offset from zero as the number of stars in the `North' and `South' bins is not always equal at each step. Over the full $360^\circ$, the offset should average to zero if the centre of the stellar sample and the pivot point of the hypothesised axis of rotation (the galaxy centre) are coincident. In this case, the green curve appears to lie mainly above the zero line, which suggests that the pivot point adopted does not coincide perfectly with the rotational centre of the sample. Changing the co-ordinates of the galactic centre in our analysis to the centre of the sample eliminates the offset, but has no significant effect on the rotation characteristics.\ 

Using a least squares method, a sinusoid of the form $\hat{y}=V_\text{rot} \sin(\theta-\theta_0)+y_0$ was fitted to the measured residual velocities, $y$, for each bin, such that $(\hat{y}-y)^2$ is minimised. The peak amplitude gives the mean rotational velocity, $V_\text{rot}$, $\theta$ tracks the \pa\ of the hypothesised axis and is stepped in $1^\circ$ intervals, and $\theta_0$  is the value of $\theta$ when $\hat{y}-y_0=0$. $y_0$ is the offset. Solving for $y_0$, $\theta_0$ and $V_\text{rot}$, $\hat{y}$ is plotted in Fig. \ref{ngc6822_motion_2}, as dotted and dashed sinusoids for the `North' and `South' bins respectively. The rotation axis \pa\ occurs at the peak of the sinusoid, so its \pa\ is given by the value of $\theta$ when $\hat{y}$ is a maximum. This occurs at $\theta_0\pm90^\circ$.\

Table \ref{rot_data} provides a summary of the results obtained, with $V_\text{rot}$ in column [1]. In column [2], we calculate the error in the velocity amplitude, $\Delta V_\text{rot}$, from the RMS of the residual velocities found by deducting the sine function from each measured value over $360^\circ$, at the best fit \pa. In columns [3] and [4], we show the \pa\ of the galactic rotation axis and an estimate of its positional error found by taking the RMS, over $360^\circ$, of $\theta-\theta^\prime$, where $\theta^\prime$ is the angle which gives the value of $\hat{y}(\theta)$ in the sine function at the best fit \pa. Column [5] shows the mean offset of the sinusoid due to the axis pivot point and the sample centre not being coincident.

\begin{table*}
\caption{Summary of sinusoidal least squares fit to mean residual velocities.}
\label{rot_data}
\centering\small
\begin{tabular}{l c c c c c}\\
&\multicolumn{2}{ c }{Rotation rate}&\multicolumn{2}{ c }{Rotation axis}&Offset\\
&$V_\text{rot}$&$\Delta V_\text{rot}$&\pa&$\Delta$\pa&$y_0$\\
&\kms&\kms&$^\circ$&$^\circ$&\kms\\
\newline
&[1]&[2]&[3]&[4]&[5]\\
\hline
&&&&&\\
30 Aug 2011&8.7&$\pm2.2$&$2^\circ$&$\pm22^\circ$&0.4\\
31 Aug 2011&11.2&$\pm2.1$&$26^\circ$&$\pm13^\circ$&0.4\\
&&&&&\\
\hline\hline
\end{tabular}
\end{table*}\normalsize

The sense of rotation on both nights is the same and is such that the south eastern side is receding and the north western side is approaching, which corresponds with the rotation of HI gas in \citet{Weldrake2003}. Moreover the rotation velocities and estimated \pa\ for each night correspond within the estimated errors which gives us good confidence in the result. Nevertheless, owing to the better quality spectra for 31 August, better \snr\ and better error results, we take the results for 31 August, i.e $V_\text{rot}=11.4 \pm2.1$ \kms\ and \pa\ $=26\pm13^\circ$ as the more reliable.\

Up to now, we have been considering only objects $<4$ kpc from the galaxy centre	. These correspond to objects in the \citet{Letarte2002} catalogue. Nevertheless there are 3 outliers at distances $\sim5$ kpc, which lie close to the stellar ellipse in Fig. \ref{ngc6822_objects} and which are classified as C-type AGBs \citep{Sibbons2015}. These objects lie in positions close to the major axis of the ellipse. If we include these outliers in our estimates, the rotation rate and rotation axis become $V_\text{rot}=11.2\pm2.5$~\kms\ $PA=-5^\circ\pm20^\circ$ for 30 August and $V_\text{rot}=12.1\pm2.3$~\kms\ $PA=19^\circ\pm14^\circ$ for 31 August, well within the margins of error for our sample.\

\begin{table*}
\caption{Comparison of results with published velocity data.}
\label{conclusions}
\centering\small
\begin{tabular}{l c c c c l}\\
&&\vh&$V_\text{rot}$&Rotation axis&Comments\\
&&\kms&\kms&$PA$&\\ \hline
This study&&$-51\pm3$&$11.2\pm2.1$&$26^\circ\pm13^\circ$&92 carbon stars\\ \hline
\citet{Koribalski2004}&[1]&$-57\pm2$&...&...&HI measurements\\
\citet{Weldrake2003}&[2]&$-53.3$ to $-54.7$&$-100$ to $+10$&...&HI measurements\\
\citet{McConnachie2012}&&$-57\pm2.0$&$47\pm2$&...&\vh\ from [1]\\
\citet{Mateo1998}&&...&$47\pm3$&...&HI measurements\\
\citet{Mateo1998}&&$-53\pm4$&...&...&Optical measurements\\
\citet{Kirby2014}&&$-54\pm1.7$&$\sim10$&...&7 red giant stars\\
\citet{Demers2006}&[3]&$+10$ to $-70$&...&$63^\circ$ to $67^\circ$&110 carbon stars\\
\citet{Veljanoski2015}&&$-59$ to $-60$&$12\pm10$&...&8 GCs, ``disk'' model, see \S \ref{s1}\\
\citet{Flores2014}&&$-57.8$&...&...&10 Planetary nebulae\\
\citet{Hwang2014}&&$-88.3\pm22.7$&....&...&4 Extended star clusters\\
\citet{Hwang2014}&&$-32.9\pm29.4$&$15.9\pm0.3$&...&Derived from carbon star population of [3]\\ \hline\hline
\end{tabular}
\end{table*}\normalsize

\section{Discussion and Conclusion}\label{S5}
Table \ref{conclusions} compares our results with the published data previously summarised in \S \ref{s1}. The heliocentric radial velocity we obtained for NGC 6822 is $-51\pm3$ \kms, based on the results from the better of the two nights of observations of a sample of its intermediate age carbon star population. This is a little lower than its HI counterpart \-- $-57\pm2$ \kms\ \-- \citep{Koribalski2004} but shows less than $1.6\sigma$ difference to this and the other estimates above, except that of \citet{Hwang2014} where it is apparent that the errors in \citet{Hwang2014} are large enough to 
include our result. In our sample, the grouping of objects in the inner and outer regions of the observed field indicates that the inner objects belong to NGC 6822, while the outer objects are more likely to be Milky Way stars, in support of the conclusion in \citet{Sibbons2015}.\

Rotation of the carbon star population about an axis having \pa\ $=26\pm13^\circ$ has been shown with a rotation speed $V_\text{rot}=11.2\pm2.1$ \kms, based on the better of the two observing nights. While we see some movement in the $PA$ of the rotation axis due to the AGB outliers in Fig. \ref{ngc6822_objects}, their numbers are too small to draw any real conclusions on its significance. Radial velocity data from more objects spread broadly throughout the stellar ellipse in Fig. \ref{ngc6822_objects} would provide better data. Nevertheless, the sense of rotation of the sample is that the SE corner is receding and the NW corner is approaching, similar to the rotation of the HI disk \citep{Weldrake2003}. We can infer from \citet{Weldrake2003}, that the \pa\ of the rotation axis of the HI disk is $\sim20^\circ$ to $\sim40^\circ$ in the plane of the sky, assuming the rotation axis is perpendicular to the major axis of the gas disk. This is comparable to our estimate of the sky projected rotation axis of the C-type stellar population in this study. Thus we find that both axes are approximately coincident and the sense of rotation of the stars and gas is the same, leading us to conclude that NGC 6822 is not a PRG.\

The \pa\ of the major axes of the isodensity contours of the RGB stars in \citet{Demers2006} varies from $80^\circ$ for the innermost contour (radius$=10^\prime$) to $65^\circ$ for the outermost contour (radius$=35^\prime$). Our inner sample has a angular radius of $\sim17^\prime$, and so falls between these extremes. Hence, there is a misalignment of $\sim45^\circ$ between our rotation axis and the respective isodensity profile. 
Furthermore, since $V_\text{rot}$ is derived from mean velocities, it is reasonable to adopt the mean distance of our carbon star sample from the galaxy centre as a basis for further comparative discussion. This distance is $\langle D\rangle=1.1$ kpc. Interpreting the velocity diagrams of \citet{Weldrake2003}, the rotation speed of the HI disk can be estimated to be $\sim30$ \kms\ at 1.1 kpc ($\approx 8$ arcmin) from the galactic centre. This rotation speed is faster than our C-star sample confirming differences in the rotation speeds of the rotationally supported gas and the pressure supported intermediate age stellar populations.\

The rotation misalignment, apparent separation of gas and stellar populations and effect of AGBs filling the gap between the central sample in this study and the stellar ellipse depicted in Fig.~\ref{ngc6822_objects} would benefit from further study, particularly at higher \snr\ and better resolving power.\

\section*{Acknowledgements}
We would like to thank Scott Croom for undertaking the AAOmega service observations and Sarah Brough for her assistance in reducing the data.\

We would also like to thank the anonymous referee for useful comments, which has helped us to write a more concise paper.





\bibliographystyle{mnras}
\bibliography{bib_MNRAS}

\newpage
\onecolumn
\appendix

\section{Tables of Data}\label{Apps}
Appendix \ref{App_A} shows the individual \fxcor\ returns for objects meeting the criteria \hght$\geqslant 0.2$, $-200$ \kms\ $\leqslant V_\text{helio} \leqslant +200$ \kms\ for 30 August 2011 and 31 August 2011. The meanings of the data in each column are described as follows:\\

\begin{tabular}{l p{14cm}}
Column&\\
1 \& 2&Fibre number and the target identifier respectively as defined by \citet{Sibbons2015}.\\
3 \& 4&Right ascension (RA) and declination (Dec) of each object (J2000) according to \citet{Sibbons2015}.\\
5&$I-$band magnitude of C-type AGBs in the inner region from the catalogue of \citet{Letarte2002}.\\
6&Distance of the object from the putative galactic centre ($\alpha=19^h:44^m:56^s$ ($296.2333^\circ$), J2000 and $\delta=-14^\circ:48':06''$ ($-14.8017^\circ$), J2000) centred on the optical co-ordinates of NGC 6822\footnote{Different catalogues and databases have differing values for the centroid of NGC 6822. These values are adopted to be consistent with \citet{Sibbons2012, Sibbons2015}} adopted by \citet{Sibbons2012, Sibbons2015} and a galactic distance of 490 kpc \citep{Mateo1998}.\\
7 \& 8&Photometric classification of each object from \citet{Sibbons2012}, and its corresponding spectroscopic classification from \citet{Sibbons2015}.\\
9 \& 13&\snr\ of each object calculated in the band 8560 \AA\ to 8650 \AA\ for the nights of 30 August 2011 and 31 August 2011 respectively.\\
10, 11 \& 12&\hght, the heliocentric radial velocity (\vh) and the velocity error ($V_\text{err}$) returned from \fxcor\ for the night of 30 August 2011.\\
14, 15 \& 16&\hght, the heliocentric radial velocity (\vh) and the velocity error ($V_\text{err}$) returned from \fxcor\ for the night of 31 August 2011.\\
17&Flag showing which objects were used in the composite templates: ``C'' for the C-type composite and ``M'' for the M-type composite.\\
\end{tabular}\normalsize

\newpage
\begin{landscape}
\centering\scriptsize
\begin{longtable}{c c c c c c c c | c c c c | c c c c | c}
\caption*{} \label{App_A}\\
\multicolumn{17}{c}{Appendix  \ref{App_A}: Individual \fxcor\ returns for objects meeting the criteria (\hght$\geqslant 0.2$, $-200$ \kms\ $\leqslant V_\text{helio} \leqslant +200$ \kms).}\\ 
\newline\\
\hline\hline
[1]&[2]&[3]&[4]&[5]&[6]&[7]&[8]&[9]&[10]&[11]&[12]&[13]&[14]&[15]&[16]&[17]\\
&&&&&&&&\multicolumn{4}{c|}{30 August 2011}&\multicolumn{4}{c|}{31 August 2011}&\\ 
\#Fibre & ID & RA & Dec & $I-$band& distance & phot. & spec. & \snr & \hght & \vh & $V_\text{err}$ & \snr & \hght & \vh & $V_\text{err}$&Flag\\
& &$(^\circ)$ & $(^\circ)$ & (mag.) & (kpc) & class. & class. & & & (\kms) & (\kms) & & & (\kms) & (\kms)&\\
\hline
\endfirsthead
\multicolumn{17}{c}{Appendix  \ref{App_A} (cont.): Individual \fxcor\ returns for objects meeting the criteria (\hght$\geqslant 0.2$, $-200$ \kms\ $\leqslant V_\text{helio} \leqslant +200$ \kms).}\\ 
\newline\\ 
\hline\hline
[1]&[2]&[3]&[4]&[5]&[6]&[7]&[8]&[9]&[10]&[11]&[12]&[13]&[14]&[15]&[16]&[17]\\
&&&&&&&&\multicolumn{4}{c|}{30 August 2011}&\multicolumn{4}{c|}{31 August 2011}&\\ 
\#Fibre & ID & RA & Dec & $I-$band & distance & phot. & spec. & \snr & \hght & \vh & $V_\text{err}$ & \snr & \hght & \vh & $V_\text{err}$&Flag\\
& &$(^\circ)$ & $(^\circ)$ & (mag.) & (kpc) & class. & class. & & & (\kms) & (\kms) & & & (\kms) & (\kms)&\\
\hline
\endhead

\hline \multicolumn{17}{r}{{Continued on next page}}\\ \hline
\endfoot

\hline\hline
\endlastfoot

1&217716&296.23&-14.86&19.3&0.52&M&Ce:&7&0.39&-10&49&9&0.48&-35&36&C\\
2&239630&296.33&-14.95&19.4&1.52&C*&C:&6&0.43&-55&41&7&0.54&-55&27&C\\
3&210316&296.43&-14.83&18.9&1.69&M&C:&5&0.59&15&25&6&0.76&-11&16&C\\
4&174035&296.37&-14.71&19.1&1.4&M*&C&3&0.37&-26&57&5&0.27&-43&67&C\\
6&199974&296.37&-14.8&19.2&1.17&M&C&...&...&...&...&9&0.65&4&24&C\\
7&194949&296.35&-14.78&18.9&1&C*&C:&12&0.72&-1&34&18&0.75&-14&23&C\\
8&242030&296.35&-14.96&18.8&1.64&...&...&6&0.6&-53&21&9&0.7&-62&22&...\\
9&220271&296.39&-14.87&...&1.43&...&...&5&0.48&-28&40&8&0.59&-63&27&...\\
10&188246&296.49&-14.76&18.3&2.12&M&C:&6&0.57&-40&29&10&0.8&-39&16&C\\
11&172656&296.54&-14.71&19.3&2.66&C*&C&5&0.29&-125&69&8&0.39&-70&41&C\\
12&211898&296.31&-14.84&18.8&0.7&M&C&9&0.57&-31&26&7&0.74&-21&20&C\\
14&246838&296.54&-14.98&...&2.99&...&...&4&0.36&-78&37&5&0.57&-61&28&...\\
15&206867&296.3&-14.82&...&0.59&...&...&7&0.42&1&58&8&0.42&-3&38&...\\
17&225812&296.32&-14.9&...&1.06&...&...&4&0.33&0&75&6&0.51&-10&31&...\\
18&190283&296.34&-14.76&19.6&0.96&M&C&6&0.45&-44&33&5&0.36&-51&36&C\\
19&168284&296.41&-14.69&19.0&1.78&C*&C:&9&0.48&40&42&10&0.48&21&34&C\\
21&213379&296.24&-14.85&19.9&0.39&C*&Ce:&4&0.4&-25&42&...&...&...&...&C\\
22&200573&296.27&-14.8&19.0&0.34&M&Ce&8&0.37&10&73&9&0.72&-12&17&C\\
23&219598&296.3&-14.87&19.0&0.79&C*&C&6&0.36&-94&77&10&0.48&-62&52&C\\
24&203106&296.34&-14.81&19.4&0.87&C*&C:&7&0.64&-31&30&9&0.68&-35&21&C\\
25&209216&296.23&-14.83&19.1&0.25&M&Ce&...&...&...&...&12&0.64&-11&21&C\\
26&197590&296.32&-14.79&19.3&0.72&C*&M:&22&0.68&-9&43&9&0.45&-28&63&...\\
28&197464&296.28&-14.79&19.0&0.43&M&C&7&0.48&-80&47&9&0.7&-58&30&C\\
29&206129&296.24&-14.82&19.2&0.16&M&dKe:&12&0.6&-32&39&12&0.54&-23&49&...\\
31&180514&296.38&-14.73&17.8&1.33&M&C&10&0.83&-40&15&11&0.89&-36&16&C\\
32&191382&296.3&-14.77&19.2&0.66&M*&C&6&0.5&0&28&8&0.51&-25&22&C\\
33&306281&297.06&-15.27&...&7.9&...&...&4&0.26&148&63&...&...&...&...&...\\
34&294767&296.99&-15.21&...&7.19&...&...&5&0.23&149&92&...&...&...&...&...\\
36&292760&296.77&-15.2&...&5.6&M&dM&23&0.57&5&50&48&0.61&7&51&M\\
38&218420&296.36&-14.87&19.0&1.21&C*&C&7&0.43&-25&26&7&0.32&-52&73&C\\
41&324220&297.01&-15.36&...&8&...&...&1&0.21&-6&26&...&...&...&...&...\\
43&229856&296.37&-14.91&19.0&1.47&C*&C&...&...&...&...&8&0.52&-31&31&C\\
45&201454&296.25&-14.8&19.4&0.15&C*&Ce:&16&0.58&-21&45&21&0.66&-45&36&C\\
47&291854&296.82&-15.2&...&5.92&...&...&...&...&...&...&5&0.3&80&51&...\\
48&184448&296.44&-14.74&19.0&1.75&M*&C&4&0.3&-30&48&7&0.31&-6&38&C\\
49&335226&296.92&-15.41&...&7.68&...&...&5&0.31&-156&63&7&0.26&2&69&...\\
51&341622&296.86&-15.44&...&7.55&...&...&4&0.23&71&106&...&...&...&...&...\\
52&342923&296.85&-15.45&...&7.52&...&...&2&0.28&33&60&...&...&...&...&...\\
53&248633&296.34&-14.99&18.9&1.82&C*&C:&6&0.46&-17&46&7&0.47&-51&38&C\\
54&229643&296.35&-14.91&18.9&1.37&C*&C&8&0.64&-55&24&10&0.71&-43&23&C\\
56&351371&296.84&-15.49&...&7.73&...&...&3&0.28&58&138&...&...&...&...&...\\
57&186954&296.47&-14.75&21.0&2&Unid*&dK&4&0.24&192&76&4&0.22&-35&65&...\\
58&249091&296.39&-14.99&...&2.08&...&...&7&0.47&-86&20&7&0.44&-38&46&...\\
62&299244&296.68&-15.24&...&5.22&...&...&...&...&...&...&4&0.28&19&62&...\\
65&220153&296.24&-14.87&19.6&0.6&M&Ce&9&0.31&-28&101&8&0.59&-2&31&C\\
68&240673&296.06&-14.96&19.7&1.93&M*&C&3&0.31&-52&66&4&0.36&-46&64&C\\
69&310311&296.6&-15.29&...&5.2&...&...&4&0.37&-178&72&6&0.33&-32&68&...\\
70&237395&296.07&-14.94&19.9&1.81&Unid*&C&3&0.26&-32&133&...&...&...&...&C\\
71&228306&296.25&-14.91&19.8&0.91&M&Ce&5&0.39&-33&61&6&0.56&-38&29&C\\
73&337540&296.5&-15.42&...&5.74&...&...&3&0.3&9&84&...&...&...&...&...\\
75&223056&296.26&-14.88&19.6&0.74&M&Ce&6&0.42&-59&55&9&0.33&-55&52&C\\
76&237999&296.09&-14.95&20.0&1.71&C*&C&4&0.24&-110&37&5&0.48&-67&27&C\\
78&220391&296.21&-14.87&...&0.63&...&...&9&0.64&-24&33&11&0.66&-26&30&...\\
80&231793&296.22&-14.92&19.0&1.04&C*&C&4&0.5&-80&57&9&0.79&-39&21&C\\
81&219171&296.23&-14.87&...&0.57&...&...&7&0.46&-79&36&8&0.72&-61&17&...\\
82&343626&296.38&-15.45&...&5.69&...&...&4&0.49&-72&24&...&...&...&...&...\\
83&343571&296.44&-15.45&...&5.81&...&...&...&...&...&...&6&0.42&71&56&...\\
84&248651&296.28&-14.99&18.9&1.65&...&...&7&0.63&-43&29&10&0.85&-56&14&...\\
85&208620&296.1&-14.83&18.5&1.12&M&C&11&0.8&-68&23&11&0.85&-66&17&C\\
86&219351&296.28&-14.87&19.4&0.7&M&Ce&7&0.52&-60&47&9&0.76&-38&22&C\\
88&226097&296.21&-14.9&19.3&0.84&M&C&5&0.53&-53&27&5&0.65&-29&21&C\\
90&215861&296.17&-14.86&19.5&0.67&M&Ce&5&0.5&-22&37&6&0.5&-13&26&C\\
91&250173&296.22&-15&19.2&1.67&Unid*&C&6&0.39&-55&58&4&0.44&-8&29&C\\
94&225648&296.16&-14.89&18.8&0.98&C*&C&11&0.84&-27&14&9&0.86&-33&16&C\\
95&242563&296.18&-14.96&19.8&1.43&C*&C&7&0.41&-41&58&9&0.64&-77&25&C\\
96&208714&296.18&-14.83&19.3&0.48&M&Ce&8&0.64&-29&22&11&0.74&-35&17&C\\
98&226603&296.12&-14.9&18.7&1.26&M&C&6&0.65&-65&24&7&0.71&-74&23&C\\
99&330842&296.34&-15.39&...&5.08&M&dM&3&0.32&-46&62&7&0.32&24&62&M\\
101&220672&296.11&-14.87&...&1.22&...&...&3&0.36&-18&41&5&0.49&-1&35&...\\
102&344297&296.12&-15.46&...&5.66&M&dM&5&0.35&85&56&12&0.52&16&46&M\\
104&209650&296.13&-14.83&19.4&0.9&M&C&6&0.37&-51&33&7&0.35&-51&39&C\\
105&233186&296.16&-14.93&20.2&1.26&Unid*&Unid&19&0.52&39&42&17&0.71&-39&26&...\\
106&344467&296.18&-15.46&...&5.61&M&dK&7&0.48&-40&33&15&0.61&-14&44&...\\
107&351774&296.12&-15.49&...&6&...&...&3&0.29&16&55&6&0.53&-62&30&...\\
112&253239&296.09&-15.01&19.4&2.14&C*&C&3&0.35&-32&26&4&0.49&-40&24&C\\
113&199754&296.21&-14.8&19.2&0.21&C*&Ce&10&0.5&-92&54&8&0.67&-45&21&C\\
115&339686&296.09&-15.43&...&5.52&...&...&4&0.31&-100&52&...&...&...&...&...\\
116&326339&295.95&-15.37&...&5.41&...&...&4&0.29&-94&75&4&0.46&24&48&...\\
119&207586&296.21&-14.82&19.1&0.26&M&C&11&0.6&-72&35&8&0.6&-61&32&C\\
122&223739&296.15&-14.89&19.8&0.98&M&C&5&0.5&-40&36&7&0.44&-4&71&C\\
123&199930&296.24&-14.8&...&0.04&...&...&14&0.78&-83&26&14&0.78&-56&23&...\\
124&239687&296.03&-14.95&19.3&2.08&M*&C&7&0.4&-15&54&5&0.65&-67&20&C\\
125&216688&296.08&-14.86&18.8&1.36&M*&C:&16&0.74&-22&18&14&0.75&-64&21&C\\
126&327377&296.01&-15.37&...&5.23&M&dK&6&0.43&-2&45&8&0.31&-9&86&...\\
128&317678&295.95&-15.33&...&5.09&M&dM:&5&0.23&-3&60&9&0.35&17&72&M\\
129&337831&295.92&-15.42&...&5.89&M&dK&4&0.37&30&36&7&0.39&36&66&...\\
130&313097&295.88&-15.31&...&5.23&M&dK&6&0.34&-62&114&12&0.48&-11&57&...\\
131&355390&295.77&-15.51&...&7.18&M&dM:&...&...&...&...&11&0.38&45&185&M\\
132&300999&295.84&-15.25&...&5.02&...&...&...&...&...&...&5&0.23&-155&63&...\\
133&372253&295.8&-15.59&...&7.68&M&dK&7&0.35&30&115&8&0.41&70&68&...\\
134&317557&295.79&-15.33&...&5.79&M&dK:&5&0.49&6&52&7&0.41&16&80&...\\
135&303607&295.81&-15.26&...&5.25&M&dM&...&...&...&...&7&0.39&26&94&M\\
136&222468&296.1&-14.88&19.5&1.32&M&C&4&0.47&-105&31&7&0.59&-28&39&C\\
137&299996&295.78&-15.24&...&5.32&M&C:&3&0.42&-46&32&5&0.53&-67&38&...\\
138&297583&295.8&-15.23&...&5.12&M&dM&8&0.44&-7&59&16&0.57&58&57&M\\
139&303122&295.74&-15.26&...&5.62&M&dM&...&...&...&...&7&0.59&15&41&M\\
141&310841&295.78&-15.3&...&5.67&...&...&...&...&...&...&8&0.46&-54&51&...\\
142&309370&295.87&-15.29&...&5.13&M&dM&8&0.37&31&63&...&...&...&...&M\\
145&350411&295.59&-15.49&...&7.93&M&dK&5&0.24&26&103&8&0.48&-51&53&...\\
147&340026&295.62&-15.43&...&7.4&...&...&4&0.23&52&77&4&0.26&-92&169&...\\
148&276398&295.72&-15.13&...&5.06&C&C:&5&0.36&-132&46&9&0.65&-111&26&...\\
149&332677&295.65&-15.4&...&7&M&dM:&5&0.35&-15&28&10&0.42&-29&50&M\\
152&336076&295.52&-15.41&...&7.89&...&...&3&0.3&-79&36&5&0.46&-39&56&...\\
153&337060&295.6&-15.42&...&7.45&...&...&3&0.26&-73&76&7&0.39&73&86&...\\
154&282194&295.69&-15.16&...&5.41&...&...&2&0.41&-27&72&4&0.39&-68&28&...\\
155&332908&295.52&-15.4&...&7.77&...&...&4&0.44&-38&31&5&0.38&11&76&...\\
156&329347&295.52&-15.38&...&7.72&M&dM:&7&0.36&12&53&12&0.55&-47&47&M\\
157&324005&295.54&-15.36&...&7.46&M&dM:&...&...&...&...&9&0.47&-82&46&M\\
158&312062&295.53&-15.3&...&7.19&M&dM:&8&0.4&-20&51&15&0.53&-16&60&M\\
159&318572&295.58&-15.33&...&7.08&M&dK&10&0.54&-19&52&15&0.61&-15&33&...\\
160&312215&295.5&-15.3&...&7.43&M&dK&...&...&...&...&10&0.31&-9&94&...\\
161&301044&295.48&-15.25&...&7.29&...&...&...&...&...&...&1&0.31&-171&36&...\\
163&298121&295.48&-15.23&...&7.24&M&dM:&...&...&...&...&8&0.38&81&68&M\\
164&305736&295.49&-15.27&...&7.32&M&dK&...&...&...&...&11&0.34&-19&89&...\\
165&271722&295.63&-15.1&...&5.63&...&...&4&0.37&-71&43&7&0.35&128&66&...\\
166&301701&295.43&-15.25&...&7.68&...&...&...&...&...&...&4&0.24&-67&77&...\\
167&277183&295.64&-15.13&...&5.65&M&dM&10&0.25&66&100&...&...&...&...&M\\
169&287590&295.45&-15.18&...&7.24&...&...&3&0.28&-156&42&6&0.49&-9&59&...\\
171&276077&295.43&-15.12&...&7.19&...&...&5&0.31&-63&36&8&0.29&15&62&...\\
172&289833&295.41&-15.19&...&7.6&M&dK&10&0.43&-17&71&15&0.73&-61&39&...\\
173&286135&295.42&-15.17&...&7.45&M&dM&7&0.33&-80&52&10&0.35&-49&60&M\\
174&245610&295.97&-14.98&19.2&2.63&M&C&3&0.33&-87&54&4&0.72&-53&28&C\\
176&248935&295.55&-14.99&...&5.91&...&...&2&0.24&-135&100&...&...&...&...&...\\
178&246831&295.64&-14.98&...&5.14&M&dK:&5&0.48&55&37&9&0.53&-9&52&...\\
180&267519&295.6&-15.08&...&5.78&...&...&2&0.3&162&69&...&...&...&...&...\\
181&254251&295.4&-15.02&...&7.09&M&dK&12&0.55&7&38&19&0.69&-2&46&...\\
183&222762&295.59&-14.88&...&5.33&M&dM&...&...&...&...&9&0.29&105&59&M\\
184&223794&295.55&-14.89&...&5.72&M&dM&10&0.31&-24&79&12&0.35&9&51&M\\
185&215764&295.52&-14.86&...&5.88&M&dMe:&5&0.31&-39&81&8&0.3&-40&54&M\\
186&223551&295.62&-14.89&...&5.11&M&dK&5&0.3&102&100&9&0.45&-52&56&...\\
187&202875&296.09&-14.81&...&1.16&...&...&4&0.4&-21&47&8&0.61&-36&37&...\\
188&239242&295.56&-14.95&...&5.71&...&...&...&...&...&...&6&0.39&-75&42&...\\
189&234263&295.54&-14.93&...&5.84&M&dM:&6&0.39&-24&32&12&0.37&-85&31&M\\
190&220386&295.54&-14.87&...&5.8&M&dK&7&0.38&-190&49&9&0.54&-110&43&...\\
191&183391&296.09&-14.74&18.5&1.31&C*&C&6&0.41&-34&41&5&0.52&-43&48&C\\
193&193714&296.11&-14.78&19.4&1.02&C*&C&7&0.42&-130&59&9&0.72&-41&26&C\\
194&184165&295.61&-14.74&...&5.18&M&dM:&5&0.41&-22&53&7&0.44&24&81&M\\
195&208174&295.51&-14.83&...&5.96&M&dMe&5&0.44&113&76&9&0.55&5&70&M\\
198&178893&296.11&-14.73&19.0&1.2&C*&C&6&0.55&-40&49&6&0.69&-42&24&C\\
199&185402&295.52&-14.75&...&5.95&M&dMe&16&0.49&58&47&20&0.66&-2&49&M\\
201&197170&296.15&-14.79&19.0&0.68&M&Ce&7&0.72&-85&20&6&0.73&-59&22&C\\
202&201578&295.53&-14.8&...&5.83&M&dMe:&9&0.45&-2&54&15&0.47&-43&34&M\\
203&216118&296.07&-14.86&...&1.41&...&...&5&0.46&-85&29&6&0.47&-108&41&...\\
204&209652&296.08&-14.83&19.3&1.26&C*&C:&...&...&...&...&6&0.44&-126&48&C\\
205&201803&295.62&-14.8&...&5.06&M&dM:&4&0.46&-122&50&5&0.3&-122&53&M\\
206&188974&296.13&-14.76&19.2&0.92&M&Ce&7&0.48&-42&33&8&0.66&-105&19&C\\
207&201131&296.16&-14.8&19.2&0.61&M&C&8&0.64&-130&25&9&0.72&-114&28&C\\
208&191195&296.11&-14.77&19.5&1.09&M&C&5&0.46&-85&27&6&0.44&-141&45&C\\
209&200928&295.56&-14.8&...&5.58&M&dMe:&6&0.32&-45&69&9&0.52&-18&42&M\\
210&201043&296.07&-14.8&19.1&1.33&M&C&4&0.38&-102&44&5&0.58&-77&21&C\\
211&205865&296.19&-14.82&19.3&0.38&C*&Ce&8&0.55&-40&41&12&0.8&-64&23&C\\
213&143954&295.41&-14.57&...&7.11&M&dM&3&0.34&-89&39&5&0.36&-37&50&M\\
214&212970&296.07&-14.84&19.5&1.43&M*&C&5&0.3&-87&88&7&0.31&-89&56&C\\
216&217770&296&-14.86&...&1.97&...&...&4&0.26&-90&44&4&0.46&-94&38&...\\
218&135229&295.57&-14.53&...&5.94&...&...&...&...&...&...&2&0.21&66&65&...\\
219&203778&296.03&-14.81&...&1.68&...&...&4&0.29&-101&94&3&0.48&-64&53&...\\
220&207096&296.16&-14.82&19.3&0.62&M&C&5&0.61&-21&25&5&0.4&-42&60&C\\
221&170279&295.97&-14.69&19.0&2.37&M&C&4&0.66&-43&24&7&0.76&-65&18&C\\
222&147856&295.59&-14.59&...&5.65&...&...&4&0.31&-54&66&4&0.45&-45&60&...\\
223&235703&296.02&-14.94&...&2.13&...&...&2&0.46&-26&61&4&0.44&-58&58&...\\
225&166319&296.11&-14.68&...&1.5&...&...&3&0.36&-101&31&5&0.46&-71&49&...\\
226&170004&296.11&-14.69&19.6&1.37&Unid*&C&5&0.34&10&77&3&0.41&-75&39&C\\
228&212351&296.17&-14.84&19.7&0.64&M&C&5&0.53&-84&42&6&0.69&-52&18&C\\
229&167638&295.97&-14.68&18.7&2.41&M*&C&6&0.5&-54&34&7&0.83&-88&16&C\\
230&174226&296.08&-14.71&19.5&1.51&M&C&2&0.43&-27&63&3&0.44&-21&33&C\\
232&100740&295.42&-14.34&...&7.82&...&...&3&0.2&-122&173&5&0.27&93&86&...\\
233&179746&296.22&-14.73&...&0.64&...&...&6&0.45&-99&31&6&0.54&-96&32&...\\
235&173378&296.1&-14.71&19.2&1.37&C*&C&5&0.46&-132&28&4&0.55&-106&40&C\\
236&92831&295.56&-14.3&...&7.03&...&...&...&...&...&...&8&0.22&-180&54&...\\
237&88989&295.46&-14.28&...&7.77&M&dM&6&0.22&-55&80&10&0.42&-143&56&M\\
238&186430&296.23&-14.75&19.0&0.44&M&Unid&13&0.46&-83&59&9&0.56&-74&31&...\\
242&78560&295.53&-14.23&...&7.63&M&dK&7&0.33&-15&76&13&0.55&-46&51&...\\
244&67411&295.58&-14.17&...&7.62&...&...&2&0.26&185&88&...&...&...&...&...\\
245&86214&295.59&-14.27&...&7.01&M&dM&10&0.38&29&49&...&...&...&...&M\\
246&191318&296.23&-14.77&18.9&0.29&C*&C&14&0.68&-76&21&10&0.79&-84&21&C\\
252&195133&296.23&-14.78&19.4&0.18&C*&Ce:&12&0.53&-63&42&10&0.6&-69&27&C\\
254&64581&295.6&-14.16&...&7.57&M&dM:&5&0.56&-41&31&9&0.55&-42&42&M\\
255&93873&295.87&-14.31&...&5.19&...&...&2&0.28&-52&67&3&0.25&-23&51&...\\
256&50172&295.62&-14.09&...&7.94&...&...&...&...&...&...&2&0.35&14&64&...\\
257&184858&296.17&-14.75&19.6&0.7&M&C&5&0.4&-38&49&7&0.6&-33&28&C\\
258&173834&296.22&-14.71&...&0.79&...&...&6&0.44&-32&50&7&0.69&-83&23&...\\
259&49090&295.74&-14.09&...&7.39&...&...&...&...&...&...&6&0.36&-137&119&...\\
260&195884&296.21&-14.78&19.5&0.22&M&C&9&0.62&-18&24&11&0.73&-57&21&C\\
261&83041&295.91&-14.25&...&5.4&...&...&4&0.38&-80&65&4&0.39&-141&59&...\\
262&165760&296.2&-14.67&20.1&1.12&M&C&6&0.44&40&37&6&0.49&-42&51&C\\
263&107892&295.81&-14.38&...&5.03&M&dK&5&0.38&-62&82&10&0.33&-82&88&...\\
266&200300&296.18&-14.8&19.7&0.41&M&Ce:&8&0.44&-44&50&8&0.53&-33&27&C\\
267&158986&296.19&-14.64&19.2&1.41&M&C&5&0.61&-61&36&5&0.5&-65&45&C\\
268&157078&296.12&-14.63&18.2&1.74&M&C:&6&0.66&-120&36&8&0.8&-88&22&C\\
269&200182&296.17&-14.8&18.9&0.52&C*&Ce&7&0.69&-37&27&9&0.67&-33&37&C\\
271&42925&295.84&-14.06&...&7.15&...&...&3&0.37&-89&67&7&0.3&-153&75&...\\
273&172060&296.2&-14.7&19.6&0.9&C*&dK&4&0.36&162&98&6&0.23&-38&96&...\\
274&168826&296.18&-14.69&19.0&1.07&M&C&5&0.67&-42&34&9&0.69&-37&29&C\\
277&76159&295.96&-14.22&...&5.5&...&...&...&...&...&...&8&0.26&-120&69&...\\
279&181364&296.13&-14.73&...&1.05&...&...&9&0.69&-54&26&9&0.64&-71&28&...\\
281&30076&295.97&-13.99&...&7.3&...&...&4&0.33&82&56&7&0.49&-23&34&...\\
282&22972&295.99&-13.95&...&7.55&...&...&4&0.42&50&36&6&0.49&76&35&...\\
283&194386&296.19&-14.78&...&0.44&...&...&10&0.76&-52&21&10&0.86&-59&15&...\\
284&173114&296.33&-14.71&18.6&1.14&M*&dM:&19&0.82&-63&22&...&...&...&...&...\\
285&78272&296.07&-14.23&...&5.09&M&dM:&...&...&...&...&7&0.48&23&47&M\\
286&27278&296.08&-13.97&...&7.19&M&dM:&...&...&...&...&9&0.39&44&89&M\\
287&179183&295.98&-14.73&19.7&2.21&C*&C&...&...&...&...&4&0.44&-53&32&C\\
289&64658&296.14&-14.16&...&5.53&...&...&4&0.42&1&47&5&0.42&46&60&...\\
290&192177&296.33&-14.77&19.7&0.87&M&C&4&0.43&-61&56&5&0.59&-51&20&C\\
291&74535&296.29&-14.21&...&5.08&...&...&3&0.26&-85&46&6&0.27&-62&47&...\\
292&169202&296.25&-14.69&19.1&0.96&M&C&5&0.57&-56&46&7&0.66&-54&15&C\\
293&159727&296.25&-14.65&...&1.34&...&...&3&0.5&-15&28&5&0.47&-57&33&...\\
294&17384&296.22&-13.92&...&7.51&...&...&...&...&...&...&5&0.38&-5&42&...\\
295&165127&296.23&-14.67&18.8&1.12&M&C&6&0.76&-52&19&8&0.68&-59&20&C\\
296&171315&296.4&-14.7&...&1.66&...&...&5&0.5&-33&37&5&0.69&-66&14&...\\
297&21034&296.25&-13.94&...&7.36&M&dK&4&0.45&70&43&7&0.65&-104&36&...\\
298&194777&296.28&-14.78&19.0&0.42&C*&Ce:&9&0.61&-72&37&9&0.72&-29&23&C\\
301&164620&296.22&-14.67&...&1.14&...&...&4&0.62&-101&29&6&0.63&-114&32&...\\
302&182253&296.31&-14.74&...&0.85&...&...&5&0.58&-69&38&5&0.62&-33&25&...\\
304&62395&296.37&-14.15&...&5.68&M&dM&6&0.46&-59&52&...&...&...&...&M\\
305&167938&296.26&-14.68&18.8&1.03&M*&C&5&0.55&-67&41&7&0.82&-76&18&C\\
306&12243&296.36&-13.9&...&7.81&M&dM:&6&0.4&21&39&8&0.37&10&46&M\\
307&20479&296.33&-13.94&...&7.43&M&dK&8&0.4&-65&53&9&0.61&-26&43&...\\
308&71601&296.34&-14.2&...&5.26&...&...&2&0.22&-35&84&4&0.33&-136&57&...\\
309&72678&296.37&-14.2&...&5.26&...&...&4&0.27&-128&87&...&...&...&...&...\\
311&61106&296.33&-14.14&...&5.69&...&...&...&...&...&...&5&0.38&81&77&...\\
313&193220&296.29&-14.77&19.1&0.53&M&Ce:&7&0.36&-86&87&5&0.46&10&32&C\\
314&198595&296.26&-14.79&19.7&0.22&M&Ce:&15&0.7&-51&28&12&0.79&-46&24&C\\
316&162531&296.28&-14.66&...&1.28&...&...&3&0.35&-7&32&4&0.44&-26&38&...\\
317&29959&296.4&-13.99&...&7.09&M&dM:&6&0.48&-11&35&9&0.51&-53&43&M\\
318&61790&296.46&-14.15&...&5.9&...&...&5&0.29&-66&53&6&0.51&-48&32&...\\
319&32087&296.55&-14&...&7.36&M&dK&5&0.29&17&76&10&0.5&88&53&...\\
320&183545&296.29&-14.74&...&0.68&...&...&9&0.58&-27&34&7&0.69&-43&28&...\\
321&196632&296.25&-14.79&...&0.16&...&...&17&0.64&-56&46&14&0.72&-117&28&...\\
322&188333&296.35&-14.76&18.6&1.05&M*&C:&27&0.74&-17&39&12&0.84&-51&19&C\\
323&174472&296.34&-14.71&19.0&1.17&C*&C&5&0.51&-57&34&7&0.57&-72&29&C\\
324&67487&296.49&-14.17&...&5.78&M&dM:&6&0.31&30&71&8&0.42&63&54&M\\
325&170941&296.29&-14.7&19.4&1&M*&dK&14&0.64&-20&42&16&0.66&-47&32&...\\
326&176863&296.35&-14.72&19.0&1.19&C*&C&7&0.53&-37&27&6&0.51&-62&23&C\\
327&83584&296.46&-14.26&...&5.03&M&dM:&9&0.34&-18&83&15&0.59&37&34&M\\
328&187649&296.24&-14.75&18.7&0.41&M&C:&9&0.66&-75&25&11&0.72&-46&18&C\\
330&167230&296.34&-14.68&20.0&1.33&C*&C&4&0.49&-31&46&7&0.51&-123&41&C\\
332&35137&296.65&-14.02&...&7.57&M&dK:&5&0.29&-27&116&8&0.27&-76&105&...\\
334&177839&296.25&-14.72&19.4&0.68&M&C&8&0.63&0&27&9&0.65&-32&28&C\\
336&38421&296.71&-14.03&...&7.68&M&dK:&...&...&...&...&6&0.28&-23&34&...\\
337&45220&296.77&-14.07&...&7.68&...&...&...&...&...&...&8&0.3&48&77&...\\
338&164010&296.33&-14.67&19.1&1.39&M&C&4&0.34&-100&66&7&0.6&-53&20&C\\
345&117821&296.71&-14.43&...&5.06&...&...&7&0.34&-40&72&...&...&...&...&...\\
348&80533&296.85&-14.24&...&7.03&...&...&4&0.37&-26&67&...&...&...&...&...\\
349&176619&296.23&-14.72&...&0.7&...&...&6&0.54&-62&34&9&0.65&-28&25&...\\
351&98241&296.65&-14.33&...&5.32&M&dM&5&0.52&-138&30&10&0.41&-42&95&M\\
352&91537&296.61&-14.3&...&5.35&...&...&3&0.44&-107&46&...&...&...&...&...\\
354&165802&296.41&-14.67&...&1.81&...&...&3&0.26&-12&82&6&0.53&-37&40&...\\
355&82795&296.95&-14.25&...&7.54&...&...&...&...&...&...&7&0.28&-141&111&...\\
357&84888&297&-14.26&...&7.81&...&...&3&0.31&-194&34&...&...&...&...&...\\
358&123455&296.83&-14.46&...&5.71&...&...&...&...&...&...&5&0.42&-48&71&...\\
359&127636&296.79&-14.49&...&5.31&M&dK:&4&0.29&-14&96&9&0.56&-15&53&...\\
360&104563&296.96&-14.36&...&7.07&...&...&3&0.43&-197&49&...&...&...&...&...\\
361&112177&296.82&-14.4&...&5.94&M&dM:&5&0.24&33&66&11&0.5&-4&75&M\\
362&81011&296.98&-14.24&...&7.81&M&dM&...&...&...&...&8&0.33&74&94&M\\
364&110475&297.01&-14.39&...&7.28&...&...&3&0.35&-5&90&5&0.4&9&64&...\\
366&152918&296.81&-14.61&...&5.06&M&C:&4&0.24&82&99&9&0.58&14&26&...\\
368&119903&296.79&-14.45&...&5.54&...&...&...&...&...&...&7&0.43&-33&85&...\\
369&167631&296.51&-14.68&...&2.49&...&...&3&0.36&-12&22&...&...&...&...&...\\
370&140011&296.8&-14.55&...&5.16&M&dM:&6&0.43&28&32&12&0.49&-9&41&M\\
371&117981&297.05&-14.44&...&7.45&...&...&...&...&...&...&8&0.4&-7&46&...\\
372&117401&297.1&-14.43&...&7.84&M&dK:&...&...&...&...&6&0.3&57&57&...\\
373&158905&296.83&-14.64&...&5.09&...&...&3&0.27&150&64&7&0.59&-22&28&...\\
374&141439&296.83&-14.56&...&5.4&...&...&2&0.49&17&40&5&0.33&18&43&...\\
375&130447&297.13&-14.5&...&7.86&...&...&2&0.27&163&55&5&0.34&11&40&...\\
376&148616&296.85&-14.59&...&5.43&M&dK:&...&...&...&...&8&0.28&5&44&...\\
377&152475&296.85&-14.61&...&5.4&...&...&...&...&...&...&10&0.26&-150&45&...\\
378&124695&297.06&-14.47&...&7.44&...&...&5&0.24&110&93&9&0.43&75&45&...\\
379&128222&297.02&-14.49&...&7.06&M&dM:&5&0.35&-40&36&9&0.25&-149&72&M\\
380&141960&297.05&-14.56&...&7.05&...&...&3&0.34&-178&53&...&...&...&...&...\\
381&136987&297.06&-14.53&...&7.21&...&...&6&0.3&-11&59&7&0.23&22&47&...\\
382&168122&296.87&-14.69&...&5.35&...&...&...&...&...&...&8&0.21&18&116&...\\
383&146932&297.07&-14.58&...&7.19&M&dM:&...&...&...&...&8&0.4&39&43&M\\
384&165837&296.92&-14.67&...&5.79&...&...&3&0.27&-96&94&...&...&...&...&...\\
388&157802&296.91&-14.64&...&5.78&M&dK:&11&0.34&7&51&...&...&...&...&...\\
389&165036&297.1&-14.67&...&7.28&...&...&...&...&...&...&7&0.25&77&136&...\\
390&155389&297.11&-14.62&...&7.42&M&dM:&...&...&...&...&14&0.27&-108&81&M\\
393&159040&297.1&-14.64&...&7.3&...&...&...&...&...&...&6&0.25&78&47&...\\
394&165107&297.15&-14.67&...&7.65&M&dM:&4&0.29&-139&53&9&0.53&12&28&M\\
395&170606&296.9&-14.7&...&5.56&...&...&...&...&...&...&5&0.33&21&58&...\\
398&204144&296.41&-14.81&19.0&1.42&Unid*&C&11&0.73&-43&25&11&0.63&-50&32&C\\
399&174894&296.85&-14.71&...&5.14&...&...&...&...&...&...&10&0.64&4&43&...\\

\end{longtable}
\end{landscape}

\normalsize

\bsp	
\label{lastpage}
\end{document}